\documentclass[usegraphicx,usenatbib]{mn2e}
      \usepackage{amsmath,amssymb,hyperref}

\def\interspace{\vspace{1.3em}}

\author[Panoglou et al.]{
  Despina Panoglou$^1$\thanks{\tt panoglou@usp.br},
  Alex~C.~Carciofi$^1$, Rodrigo~G.~Vieira$^1$,
  Isabelle~H.~Cyr$^2$, \and~Carol~E.~Jones$^2$,
  Atsuo T.~Okazaki$^3$, Thomas Rivinius$^4$
      \\\\
  $^1$Instituto de Astronomia, Geof\'isica e Ci\^encias Atmosf\'ericas, Universidade de S\~ao Paulo, SP 05508-900, Brazil\\
  $^2$Department of Physics and Astronomy, University of Western Ontario, London, Ontario N6A 3K7, Canada\\ 
  $^3$Faculty of Engineering, Hokkai-Gakunen University, Sapporo, Hokkaido 062-8605, Japan\\ 
  $^4$European Organisation for Astronomical Research in the Southern Hemisphere, Casilla, Santiago 19001, Chile}
\title[Be discs in binary systems]{Be discs in binary systems. I.~Coplanar orbits}

    \def\oned{one-dimensional}
    \def\twod{two-dimensional}
    \def\threed{three-dimensional}
    \def\SPH{smoothed particle hydrodynamics}
    \def\VDD{viscous decretion disc}
\newcommand{\Msun}{M_\odot}
\newcommand{\Minj}{\dot{M}_\textrm{inj}}
\newcommand{\Mdec}{\dot{M}_\textrm{dec}}
\newcommand{\Srad}{\left<\Sigma_p(r)\right>_\phi}
\newcommand{\Sradz}{\left<\Sigma_{p=0}(r)\right>_\phi}
\newcommand{\Sone}{\Sigma_0}
\newcommand{\Sonep}{\Sigma_{0,p}(\phi)}
\newcommand{\Rtpp}{R_{t,p}(\phi)}
\newcommand{\Rtp}{R_{t,p}}
\newcommand{\ass}{\alpha_\text{ss}}
\newcommand{\Porb}{P_\text{orb}}
\newcommand{\Tav}{T_\text{av}}
\newcommand{\Td}{T_\text{av}}
\newcommand{\Teff}{T_\text{eff}}
\newcommand{\Rbar}{\overline{R}_p(\phi)}
\newcommand{\RL}{R_\text{L}}
\newcommand{\RLs}{R_\text{L,s}}
\newcommand{\RLBe}{R_\text{L,Be}}
\newcommand{\tauobs}{\tau_\text{obs}}
\newcommand{\phiel}{\phi_\text{el}}

\date{Accepted 2016 June 20. Received 2016 May 20; in original form 2015 December 10}

\begin{document} 
\maketitle
\begin{abstract}
{Be stars are surrounded by outflowing circumstellar matter structured in the form of \emph{decretion discs}. They are often members of binary systems, where it is expected that the decretion disc interacts both radiatively and gravitationally with the companion.}
{In this work we study how various orbital (period, mass ratio and eccentricity) and disc (viscosity) parameters affect the disc structure in coplanar systems.}
{We simulate such binaries with the use of a smoothed particle hydrodynamics code.}
{The main effects of the secondary on the disc are its truncation and the accumulation of material inwards of truncation.
We find two cases with respect to the effects of eccentricity: (i) In circular or nearly circular prograde orbits, the disc maintains a rotating, constant in shape, configuration, which is locked to the orbital phase. The disc is smaller in size, more elongated and more massive for low viscosity parameter, small orbital separation and/or high mass ratio.
(ii) Highly eccentric orbits are more complex, with the disc structure and total mass strongly dependent on the orbital phase and the distance to the secondary.
We also study the effects of binarity in the disc continuum emission. Since the infrared and radio SED are sensitive to the disc size and density slope, the truncation and matter accumulation result in considerable modifications in the emergent spectrum.}
{We conclude that binarity can serve as an explanation for the variability exhibited in observations of Be stars, and that our model can be used to detect invisible companions.}
\end{abstract}

\section{Introduction}\label{s:intro}
Be stars are rapidly rotating main-sequence or giant stars of spectral type between late O and early A. They are surrounded by decretion discs, i.e.~an outflowing circumstellar structure formed by material ejected by the star. 

Most massive stars ($\simeq71\%$) are components of a binary system, or have existed as such during some period in their evolution \citep{Sana12}. A substantial fraction of Be stars seem to follow this trend \citep[1/3 of the galactic Be stars are in binary systems according to][]{OuPa10}, with their decretion discs very often having the role of the reservoir from which the secondary accretes matter.

The Be disc may interact radiatively and/or tidally with the companion star, depending on its nature: The companion is often a compact star emitting X-rays, a hot subdwarf O or B star \citep[sdO/sdB star, see][]{Heb09} directly observable by the UV radiation it emits, or a low-mass main-sequence star, which is the most difficult case to detect.
High energy interaction occurs in Be/X binaries when the compact companion produces X-ray luminosity as it accretes some of the matter ejected through the Be disc \citep{Reig11}. Be/pulsar binaries may exhibit $\gamma$ radiation, when the particles accelerate in the shocked region between the pulsar wind and the Be disc or wind \citep{SiBe05a,TaOk12}.
The two stars in Be binary systems can also interact radiatively, especially when the much hotter (usually sdO) star irradiates the Be disc, rising its temperature and affecting the strength of recombination lines \citep[e.g.~Balmer lines,][]{GBFK98}; through such spectral features has it been made possible to detect Be/sdO binaries. Finally, the tidal interaction between the disc and the companion affects the disc structure, causing disc truncation and/or warping, as shown in the model of \cite{OkBa02}.
The density structure of the disc of outflowing gas surrounding binary Be stars is the main focus of this work.

Various theories have been proposed to predict the observed properties of Be discs, but observational constraints have limited the possible candidates to one \citep{RiCa13}: the {\VDD} (VDD) model proposed by \cite{LeOs91}.
The physics of viscous decretion through a disc is the same as in viscous accretion in compact stars \citep{Prin81}, with the only difference being the radial direction of flow.
\citeauthor{Port99} (\citeyear{Port99}; see also \citealt{Okaz01}, \citealt{BjCa05}) found that the {\oned} surface density $\Sigma(r)$ of viscous decretion discs of isolated stars is a function of the cylindrical distance $r$ from the central star, under the assumption of an isothermal keplerian rotating disc fed at constant rate for an adequately long period of time, and is approximately given by
  \begin{equation}
  \Sigma(r) = \Sigma_0 \left(\frac{r}{R_*}\right) ^{-2} \label{e:sd},
  \end{equation}
where $R_*$ is the equatorial radius of the star. This simple model has been successfully applied to studies of individual stars \citep[e.g.][]{CMKB06,JoTy08} and statistical studies \citep[e.g.][]{ToGi11,SJTS10}. For a recent review on the status of the VDD model see \cite{RiCa13}.

A steady-state model is not appropriate for most Be stars due to their intrinsic variability, exhibited in all types of observables.
The variability of Be stars may be both long- and short-term, and stems mainly from variations in the structure of the decretion disc, which is mostly associated with the disc feeding process. In some cases the disc completely disappears, and then reappears in a series of outbursts \citep[see][for a well documented case]{BjMi02}. The Be stars also show variability in their emission line profiles, e.g.~$V/R$ variations \citep{Okaz91}, short-term absorption features \citep{SmPo93}, line wing enhancements \citep{SmCo97}, and Be/shell phase transitions \citep{DMMP86}.

In recent years some effort has been made in developing dynamical models of Be stars to verify whether the VDD model offers a viable explanation for the temporal variability of their discs. \cite{HCRO12} studied the dynamical evolution of a disc of an isolated Be star fed with matter ejected from its surface.
Modifying the disc feeding rate from continuous to periodic or episodic, and applying one-dimensional hydrodynamical calculations to describe the viscous diffusion of the disc material, the authors managed to qualitatively reproduce the long-term photometric variations. \cite{CBOO12} successfully fitted the light curve of 28 Canis Majoris (28 CMa) during an event of disc dissipation beginning in 2003 that was observed and characterised by \citet{SBRO03}.

Not all variability, however, can be attributed to variations of the disc feeding rate and/or the process of viscous diffusion throughout the disc.
For instance, the long-term high-amplitude $V/R$ variations are most probably due to global density waves within the disc \citep{Okaz91,CaOk09,EsCa15}.
Shorter periods and smaller-scale $V/R$ variations (usually veiled when the decretion disc is too massive or due to the much stronger density wave $V/R$ variations) are thought to be caused by either asymmetric mass ejections from the star, or driven by a binary companion, in which case they would result in disc deformation due to tidal interaction. For example, the phase-locked variations in the binary system of $\pi$ Aquarii are thought to be connected with binarity since the $V/R$ ratio varies with the orbital phase \citep{BjMi02,Poll12a}.

The majority of VDD studies in the literature are focused on the disc properties around single stars. An exception to this is the work of Okazaki and his collaborators \citep{OkBa02,NeOk01a,OkNe01a,RoOO07}, who explored the tidal interaction in Be/X-ray binaries assuming that the disc is isothermal and the two stars have coplanar prograde orbits.
To this aim, \cite{OkBa02} used a {\threed} {\SPH} (SPH) code, in order to solve the time-dependent hydrodynamical equations for a flow of an ensamble of particles of negligible mass that are ejected axisymmetrically from the stellar equator.

In this work, we take advantage of the technological progress achieved since that publication, allowing for faster simulations. Our main goal is to extend the work of \cite{OkBa02}, in order to, on one hand, understand how various orbital and disc parameters affect the disc structure and, on the other hand, predict what kind of observational signatures we should expect depending on the values of those parameters.
We will thus explore the parameter space in terms of different values of viscosity, mass ratio and separation between the two stars. In this first paper we examine the cases of coplanar circular and eccentric orbits, both prograde and retrograde.
Misaligned discs, i.e.~discs tilted at some angle with respect to the orbital plane, are examined in a follow-up publication \citep{CyJo15}.

Section \ref{s:model} provides a description of our model. It is followed by a presentation of the results for our reference simulation (Section \ref{s:ref}).
The remainder of the text is separated into three main parts: circular (Section \ref{s:procrc}) and elliptical (Section \ref{s:proell}) prograde orbits, and retrograde orbits (Section \ref{s:rtrcrc}). The observational effects of the binary interaction on the disc continuum emission are examined in Section \ref{s:pseudo}. A summary and discussion is given in Section \ref{s:end}.

\section{Model description}\label{s:model}
We use the same {\threed} SPH code as the one used by \citet[see also \citealt{BaBo97}]{OkBa02}.
The ejection of mass from the Be star is modelled by constant and axisymmetric ejection of particles of keplerian rotational velocity just outside the equatorial radius. Mass is injected to the decretion disc at a distance $r_\textrm{inj}=1.04R_*$ and at a rate $\Minj$, constant in time.

The overall decretion rate is much smaller: As soon as each particle is injected into the disc, it interacts with the existing particles. From this interaction most of the injected particles lose their angular momentum and are re-accreted onto the Be star. Only a few particles manage to escape outwards, retaining the angular momentum of the particles with which they interacted as soon as they were injected.
The ratio between the mass injection rate and the mass that actually flows outwards, $\Mdec$, decreases quickly in the beginning of the simulation and reaches an asymptotic value as the mass of the disc grows \citep[see fig.~5 of][]{OkBa02}.
Our simulations result in a decretion rate of $\Mdec<0.1\Minj$.

The accretion boundary of the secondary has a variable radius depending on its Roche lobe, given by $R_\textrm{acc}=\RLs$.
The Roche lobe radius is estimated with the Eggleton formula \citep[][Eq.~2]{Eggl83}. \citeauthor{Eggl83} gave an approximation for the effective radius $\RL$ of the Roche lobe of a binary star, i.e.~the region within which any orbiting material is bound to the star, as a function of the orbital separation $a$ and the binary mass ratio $q_r=M_x/M_y$ (where $M_y$ is the mass of the primary star and $M_x$ is the mass of the secondary, if $\RL=\RLs$ refers to the secondary):
          \begin{equation}
          \frac{\RL}{a} = \frac{0.49q_r^{2/3}}{0.6q_r^{2/3}+\ln(1+q_r^{1/3})}\quad,\quad
          q_r<1. \label{e:Egg}
          \end{equation}
\cite{OkBa02} substituted the separation $a$ with the instantaneous distance $D$ in this relation, in order to account for eccentric binaries (especially for highly eccentric binaries where the distance between the two stars at periastron becomes very small).
When this relation is used to evaluate the Roche lobe radius of the primary, $\RL=\RLBe$ and the mass ratio is reversed.
This estimation for the accretion radius serves adequately the purposes of this work, since a smaller accretion radius of the secondary star causes the simulations to be highly time-consuming. In any case, we are not interested in the accreting star, but only in the decreting Be star and the properties of the disc.

In the past, simulations of the binary orbits of Be/\mbox{X-ray} binary systems typically assumed that the Be decretion disc is big enough so that the compact object accretes mass from the decretion disc near the periastron.
\cite{OkNe01a,OkNe01b} constructed a semi-analytical model that shows that the disc in such systems is truncated at a distance smaller than the periastron separation as long as $\ass\ll1$, where $\ass$ is the Shakura-Sunyaev viscosity parameter \citep{ShSu73}.
The model of \cite{OkNe01a,OkNe01b} gives an estimation of this distance, which we shall call the \emph{truncation radius} $R_t$, by comparison of the resonant torque with the viscous torque.

The viscous torque and the tidal torque tend to rotate the gas around an axis perpendicular to the plane of flow.
The viscous torque tends to rotate in the direction of the outward spiralling gas flow, while the tidal torque tends to slow down the rotation of the gas. Both work in opposite directions, but the tidal torque is too weak to match the viscous torque. Indeed, it is the resonant torque that causes truncation, at the smallest resonant radius where the resonant torque exceeds the viscous torque.

The results by \citet{OkBa02} indicate that the so-called disc truncation does not simply interrupt the disc in radial distances closer to the star compared to the isolated case, but it causes a sudden drop in the density of the gas. In practice, this means that the structure of the decretion disc in Be binaries is not well represented by a single power law as given by Eq.~\eqref{e:sd}.

\subsection{Timescale considerations}\label{s:teev}
\cite{HCRO12} found that at constant feeding rates, the disc grows faster for increased values of $\ass$. Starting from a disc-less state, once the disc feeding is turned on, the disc grows inside out, with the inner regions quickly reaching a quasi-steady state and the outer disc filling up at slower rates (fig.~2 of \citeauthor{HCRO12}).

When studying the temporal evolution of a Be disc in a binary system, the surface density profile can be better studied when comparing it at different points in time but at the same orbital phase $p$, with $p\in[0,1)$, $p=0$ corresponding to the apastron, and $p=0.5$ to the periastron. In a circular system, $p=0$ simply means that the secondary has orbited the Be star for a time equal to an integer number of periods since the beginning of the simulation at a disc-less state.
Within this context, the evolution time of the system can be expressed as $t=(k+p)\Porb$, where $\Porb$ is the binary orbital period, and $k$ is an integer corresponding to the number of whole periods that have passed since the disc started forming. It now becomes clear that two time-scales control the state of a disc in a binary system: the orbital period and the viscous diffusion time-scales.
As we shall see below, the interplay of these two time-scales can cause complex surface density profiles, particularly in eccentric binaries.

This periodicity in the density structure makes the use of the term `quasi-steady' state necessary, since even at later stages of the disc evolution the surface density profile never remains constant in time throughout one orbital period.
In order to decide whether the system has reached a quasi-steady state, we cannot compare subsequent time steps, but, instead, same phases of subsequent cycles.
From here on, we will use the terms `steady-state' and `quasi-steady state' interchangeably, with both terms referring to an adequately evolved disc configuration.

\section{Reference system}\label{s:ref}

We adopt a reference binary system that corresponds to a star with the parameter values for $\zeta$ Tauri as listed in \cite{RBHF09}, with the exception of the orbital period which was set to 30 days instead of $\zeta$ Tauri's actual period of 133 days, as tidal effects are much stronger for closer binaries. In particular, our reference run is for a binary system consisting of a Be star with mass $M_1=11.2\,M_\odot$ and radius $R_*=5.5\,R_\odot$.
The binary parameters are: mass ratio $q_r=0.08$, semi-major axis $a=17\,R_*$, eccentricity $\epsilon=0$. In addition, the reference system is prograde, in the sense that the star, the disc and the companion, all rotate in the same direction. 
The disc is considered isothermal and has a fixed average temperature equal to $\Tav=0.6T_\text{eff}$ \citep{CaBj06} with $T_\text{eff}=19\,400$~K, while the Shakura-Sunyaev viscosity parameter of the disc is equal to $\alpha_\text{ss}=0.4$ \citep{CaOk09}.

Each particle has a mass chosen to be of the order of $10^{-14}M_\odot$. The final number of particles is of the order of $10^4$, and the injection rate is kept constant and equal to \mbox{$\Minj=10^{-7}M_\odot/$yr}.

\begin{table*}\centering
\caption{Summary of selected simulations. The injection rate $\Minj$ remains the same, and the total mass injected per time step is divided in 5\,000 particles in all simulations. The run time is such that it is always past the quasi-steady state. As all simulations (except from the first one) in this table have the same period (30 days), the time unit for the time to steady state and the simulated time is also 1 month.
Listed are also the number of particles $N_\text{sph}$, the decretion rate $\Mdec$ and the total disc mass $M_\text{disc}$ at the end of each simulation.
The last column refers to the figure of the temporal evolution and/or a figure with snapshots of the simulation within the text, if there are any.}\label{t:chars}
\begin{tabular}{rccccccrccc}
       \hline
$\Porb$&$q_r$&$\ass$&$\epsilon$& orbital  &time to steady state&run time
              &$N_\text{sph}$&$\Mdec$&$M_\text{disc}$ & figures\\
       &      &      &         & direction&($\Porb$)           &($\Porb$)
              &  &($10^{-9}\Msun/$yr)&($10^{-9}\Msun$)\\
       \hline
       \hline
133 & 0.08 & 0.4 & 0.0 &   prograde & 24 & 25.0 & 13\,327 & 0.701 & 3.865 &\ref{f:tempo}c         \\ 
 30 & 1.00 & 0.4 & 0.0 &   prograde & 55 & 84.0 & 58\,574 & 0.895 & 3.832 &\ref{f:tempo}d         \\ 
 30 & 0.08 & 1.0 & 0.0 &   prograde & 21 & 30.0 & 21\,822 & 1.085 & 1.428 &\ref{f:tempo}b\\
 30 & 0.08 & 0.4 & 0.0 &   prograde & 42 & 51.5 & 46\,361 & 1.195 & 3.033 &\ref{f:tempo}a, \ref{f:aaec1}\\
 30 & 0.08 & 0.4 & 0.6 &   prograde & 18 & 30.0 & 27\,009 & 1.112 & 1.767\\ 
 30 & 0.08 & 0.4 & 0.9 &   prograde & 13 & 30.0 & 10\,513 & 3.051 & 0.688 &\ref{f:tempo}e, \ref{f:aaec2}\\
 30 & 0.08 & 0.4 & 0.0 & retrograde & 27 & 30.0 & 38\,488 & 1.562 & 2.518 &\ref{f:tempo}f, \ref{f:daaac}\\
 30 & 0.08 & 0.4 & 0.9 & retrograde & 21 & 22.0 &  9\,384 & 2.562 & 0.614 & \ref{f:dretr}\\
       \hline
\end{tabular}
\end{table*}

\begin{figure}\centering
\includegraphics[scale=.17]{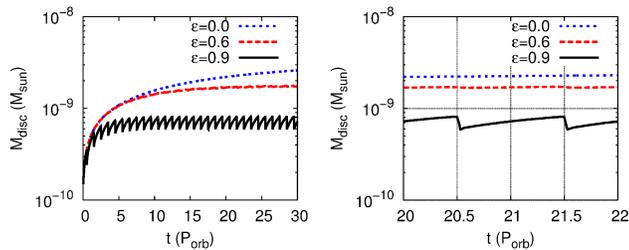}
\caption{The disc mass as a function of time for prograde systems of three different eccentricities. The left panel shows the evolution since the beginning of the simulation, and the right panel zooms in for time between 20-22 $\Porb$.}
\label{f:Mdisc}
\end{figure}

Along with the reference system which will be examined in the current section, we performed a series of simulations with differing parameters ($\ass$, $\Porb$, $q_r$, $\epsilon$, orbital direction). Their effect on the disc density will be discussed further on (Sections \ref{s:procrc}-\ref{s:rtrcrc}).
An overview of the characteristics for selected simulations is given in Table~\ref{t:chars}, which gives information both about the time needed for each system to reach the steady state, as well as the total simulated time.
The final number of particles $N_\text{sph}$, decretion rate $\Mdec$ and total disc mass $M_\text{disc}$ are also indicated. The decretion rate $\Mdec$ is calculated as the asymptotic value reached past the steady state.
More specifically, it is estimated as the fitting to a constant value of the last timely SPH data (after the steady-state). The same method for the asymptotic values is followed for the final number of particles and the disc mass in the highly eccentric cases ($\epsilon>0.6$), where these two quantities oscillate around a mean value, depending on the position of the secondary (see Fig.~\ref{f:Mdisc}).

\subsection{Azimuthally averaged properties}
\label{s:best}
As explained in Section \ref{s:teev}, in a sufficiently evolved binary system the disc structure depends on the orbital phase $p$. Therefore the azimuthally averaged surface density has to be defined with respect to the orbital phase $p$ it refers to:
\begin{equation}
\Srad = \frac{1}{2\pi} \int_{0}^{2\pi} \Sigma_p(r,\phi)~\text{d}\phi
\end{equation}
where $\phi$ is the azimuthal angle, and $\Sigma_p(r,\phi)$ is the surface density.
In order to describe the global properties of the disc, we will use the azimuthally averaged quantity $\Srad$ at a time equal to integer number of orbital periods since the beginning of the simulation, $p=0$. As shown by \citet[][eq.~27]{BjCa05}, the surface density scales with $\ass$ and $\Mdec$ as \mbox{$\Sigma\propto\Mdec\ass^{-1}$}. We therefore plot $\ass\Srad$ to enable comparison between simulations of different viscosities.

\begin{figure}\centering
\includegraphics[width=\linewidth]{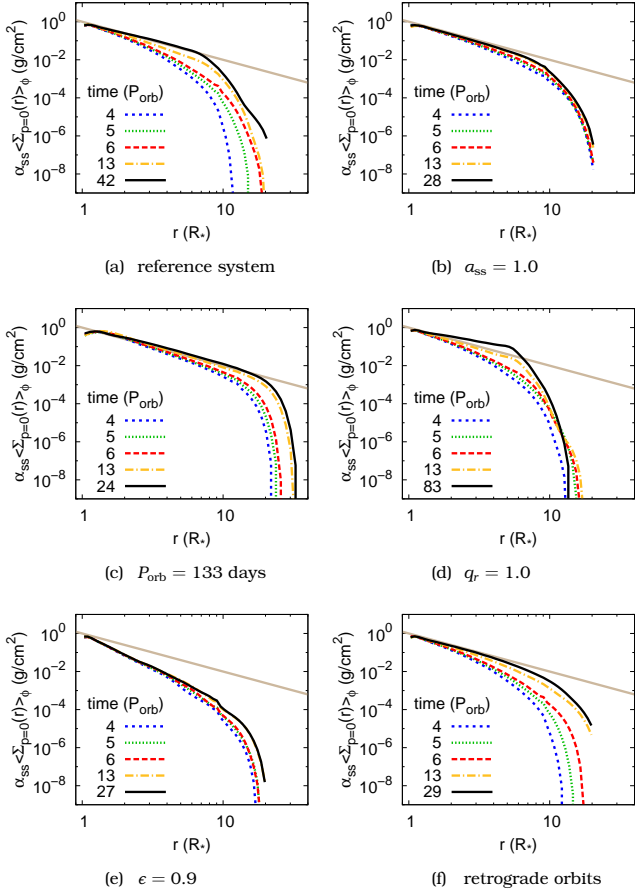}
\caption{Azimuthally averaged surface density at $p=0$ for the reference system (panel a; $\ass=0.4$, $\Porb=30$~days, $q_r=0.08$, \mbox{$\epsilon=0$}, prograde orbit) and for five other simulations differing to the reference system only with respect to the value of one parameter: (b) $\ass=1$, (c) $\Porb=133$ days, (d) $q_r=1$, (e)~$\epsilon=0.9$, (f)~direction of rotation of the binary motion with respect to the disc rotation (retrograde orbit).
Each panel plots the surface density at the periastron passage at five different points in time; from bottom up more evolved, the first four being $t=4,5,6,13~\Porb$ -- or $t=4,5,6,13$ months for all other systems except for (c), for which $t=18,22,27,58$ months; the fifth time point corresponds to a point past the quasi-steady state -- different for each system. The solid straight line shows the density profile of Eq.~\eqref{e:sd}.}
\label{f:tempo}
\end{figure}

Figure \ref{f:tempo}a shows the temporal evolution of the surface density profile of our reference system for a few selected epochs. The quasi-steady state is reached in 42 orbital cycles (1\,260 days or 3.5 years; see Table \ref{t:chars}).
The initial evolution, before signs of truncation, is similar to that of an isolated star, as exhibited in the simulations of \cite{HCRO12}. However, as the disc grows in mass, the influence of the resonant torque from the companion becomes more evident. The density structure evolves towards a profile in which the inner part approximately follows Eq.~\eqref{e:sd}, and the density fall-off of the outer part is much steeper with the radial distance than that of the inner disc.

\begin{figure}\centering
\includegraphics[scale=.2]{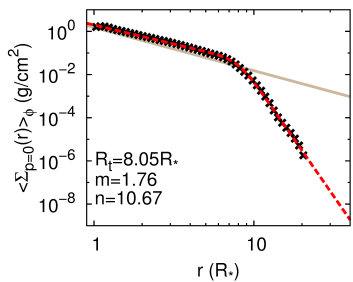}
\caption{Azimuthally averaged surface density in quasi-steady state (crosses). The fit of Eq.~\eqref{e:Atsuo} is the dashed (red) line; the straight line is the density profile of Eq.~\eqref{e:sd}. The values of the fitted parameters are denoted.}
\label{f:aarpl}
\end{figure}

The surface density was fitted using the following equation \citep{OkBa02}:
          \begin{equation}
          \Srad = A\frac{(r/R_t)^{-m}}{1+(r/R_t)^{n-m}}\label{e:Atsuo}
          \end{equation}
where $R_t$ corresponds to the truncation radius, $m$ is the power-law exponent of the inner region of the disc, and $n$ is the outer exponent ($\lim_{r\rightarrow\infty}\Srad\propto\lim_{r\rightarrow\infty}r^{-n}=0$).
Equation \eqref{e:Atsuo} acknowledges the fact that truncation changes the density distribution with the radial distance, thus it is an appropriate description of the numerical results for the surface density.
The results of fitting the steady-state surface density for our reference system with Eq.~\eqref{e:Atsuo} are shown in Fig.~\ref{f:aarpl}.

\subsection{Phase dependence and azimuthal structure}
\label{s:phase_dependent}
\begin{figure}\centering
\includegraphics[width=\linewidth]{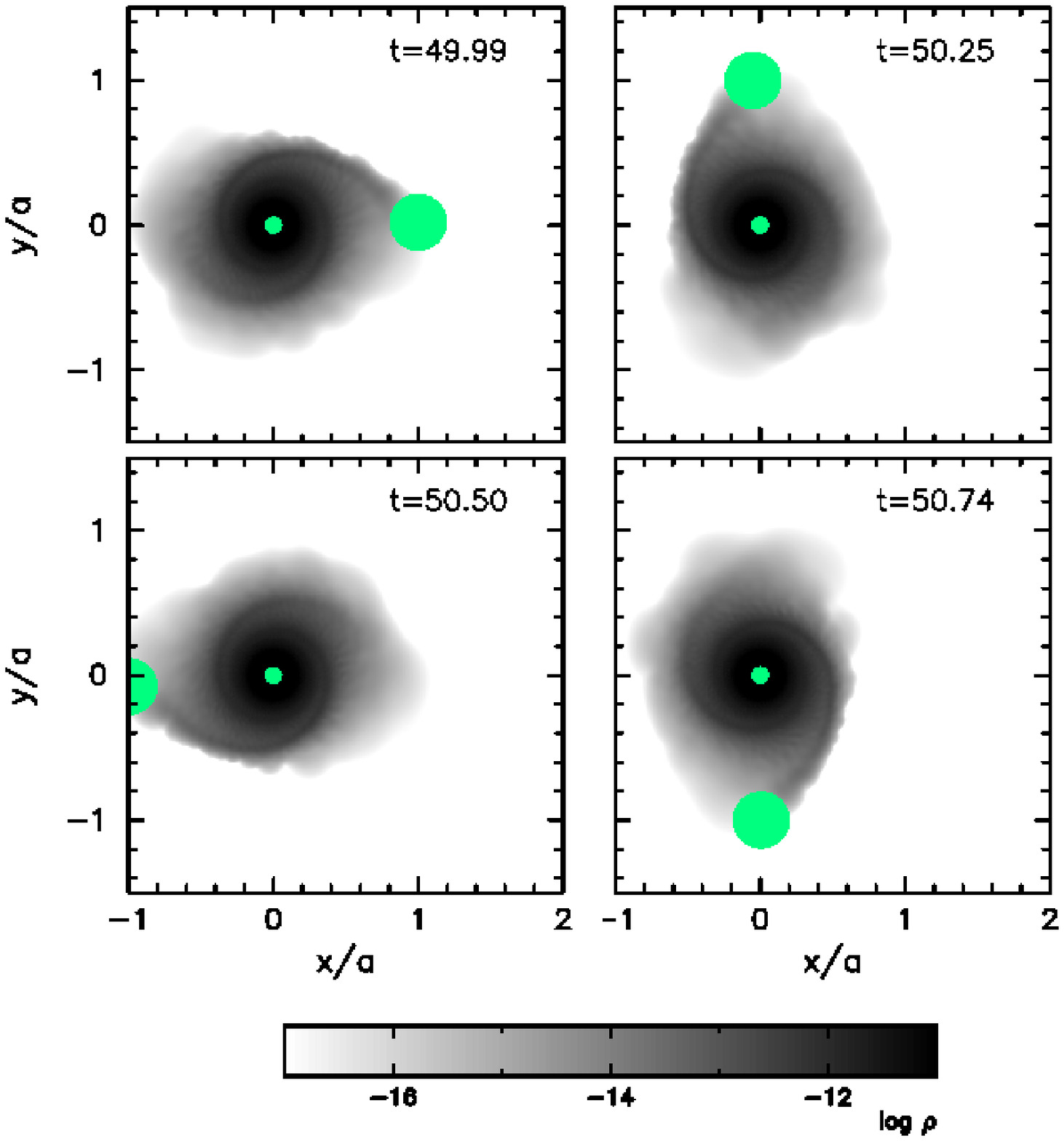}
\caption{Snapshots of the reference binary system at quasi-steady state. The distances are given in semi-major axis units.}
\label{f:aaec1}
\interspace
\includegraphics[clip,trim=0cm 1cm 0cm 1cm,scale=.17]{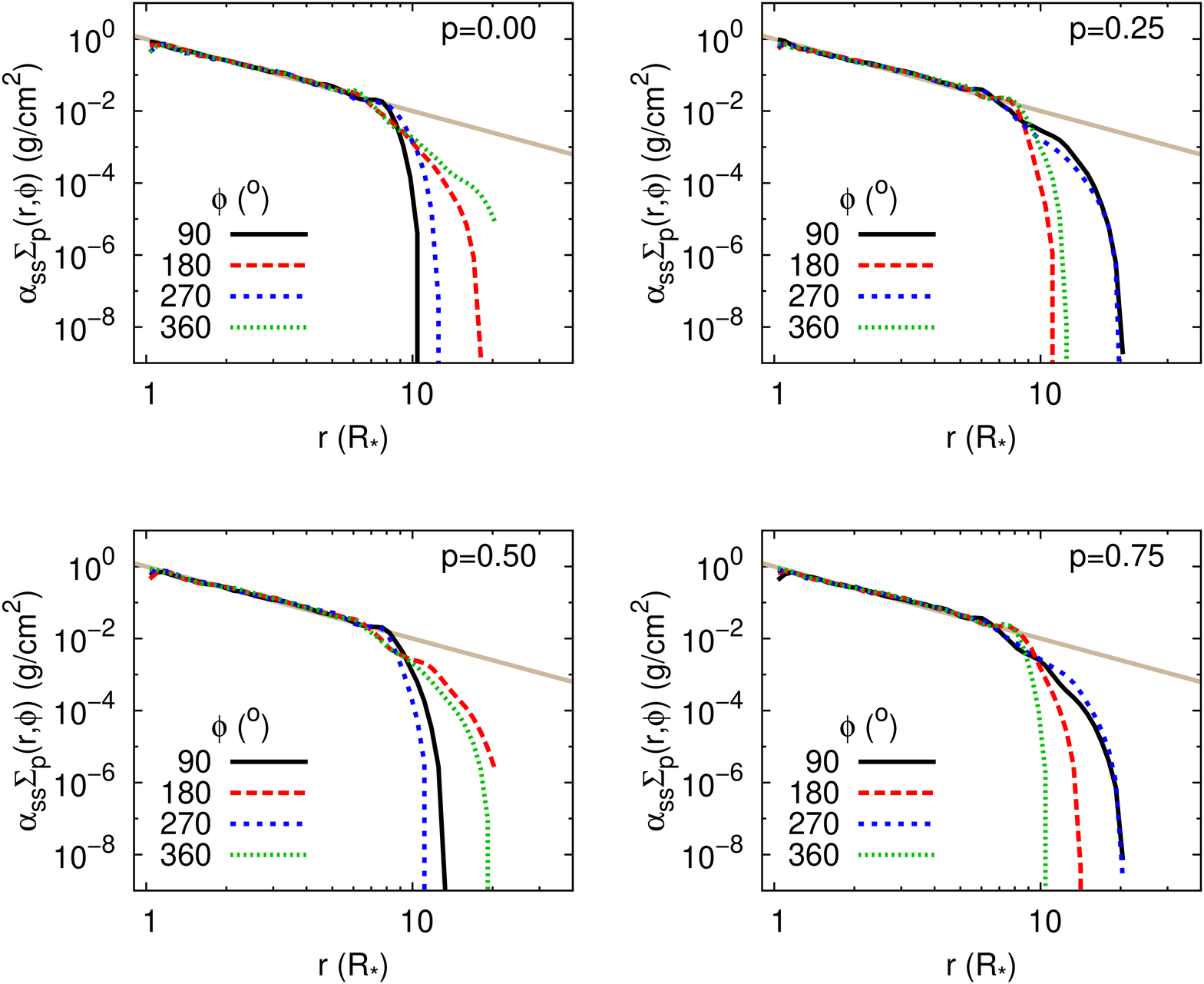}
\caption{Surface density for the reference system as a function of the radial distance from the centre of the star at four azimuthal angles. Each panel shows steady-state results for the indicated phase, $p$. The straight line is the density profile of Eq.~(\ref{e:sd}).}
\label{f:cont41}
\end{figure}

In Fig.~\ref{f:aaec1} we show snapshots of the {\twod} density structure in four orbital phases ($p=0.00$, 0.25, 0.50, and 0.75), with all figures centered at the primary. The positions of the two stars are marked with circles. The size of the primary is given by its radius, while the secondary has the size of its accretion radius estimated with the Eggleton formula.
In Fig.~\ref{f:cont41} the surface density profile is plotted for four values ($\phi=90,180,270,360\degr$); each panel shows the disc structure at a different orbital phase. When the truncation radius is minimum (for $\phi=180,360${\degr} when $p=0$ and 0.5), the outer density slope is also smaller, and vice versa.
In general, we can see that the inner part of the disc is largely axisymmetric showing little dependence on $\phi$, while its extent as well as the density slope of the outer part changes with the azimuthal angle $\phi$.

To draw more quantitative information about the {\twod} structure of the surface density, we divided the azimuthal angle range $[0,2\pi]$ into 25 intervals and fitted the radial surface density profile with Eq.~\eqref{e:Atsuo} for the mean angle of each interval.
This was repeated for the four orbital phases displayed in Fig.~\ref{f:aaec1}. In order to reduce the noise, the same process was averaged over five consequent orbital cycles after the system has reached the steady state. We thus obtained functions of $\phi$ for the truncation radius and the density slopes at each phase, that allows us reconsider Eq.~\eqref{e:Atsuo} with its parameters as functions of $p$ and $\phi$:
          \begin{equation}
	  \Sigma_p(r,\phi)
          = A_p(\phi)\frac{(r/R_{t,p}(\phi))^{-m_p(\phi)}}
	  {1+(r/R_{t,p}(\phi))^{n_p(\phi)-m_p(\phi)}}.\label{e:Atsne}
          \end{equation}

\begin{figure}\centering
\includegraphics[scale=.17]{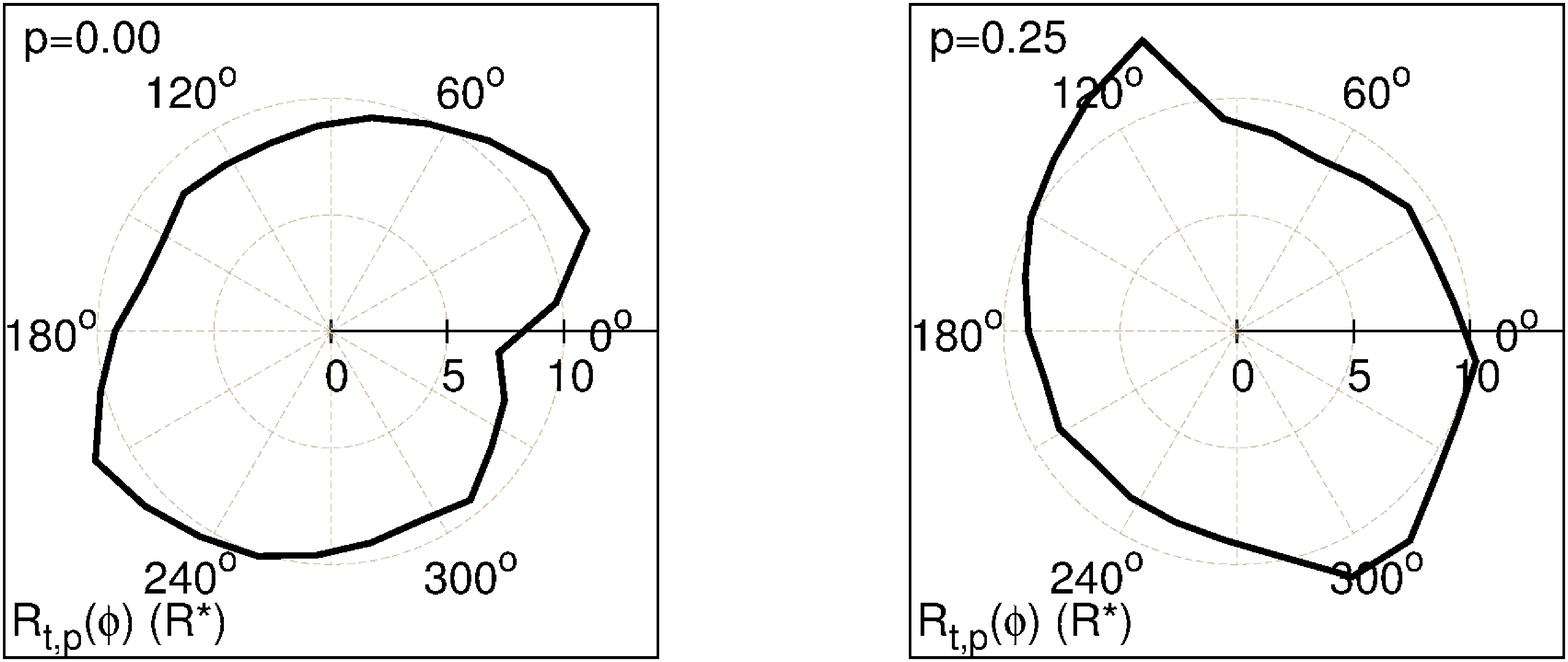}
\caption{Polar plots of the truncation radius, obtained by fitting the surface density of two indicative snapshots from Fig.~\ref{f:aaec1} with Eq.~\eqref{e:Atsne} for 25 values of $\phi$. The Be star lies at the centre of each plot.}
\label{f:41-16}
\interspace
\includegraphics[scale=.17]{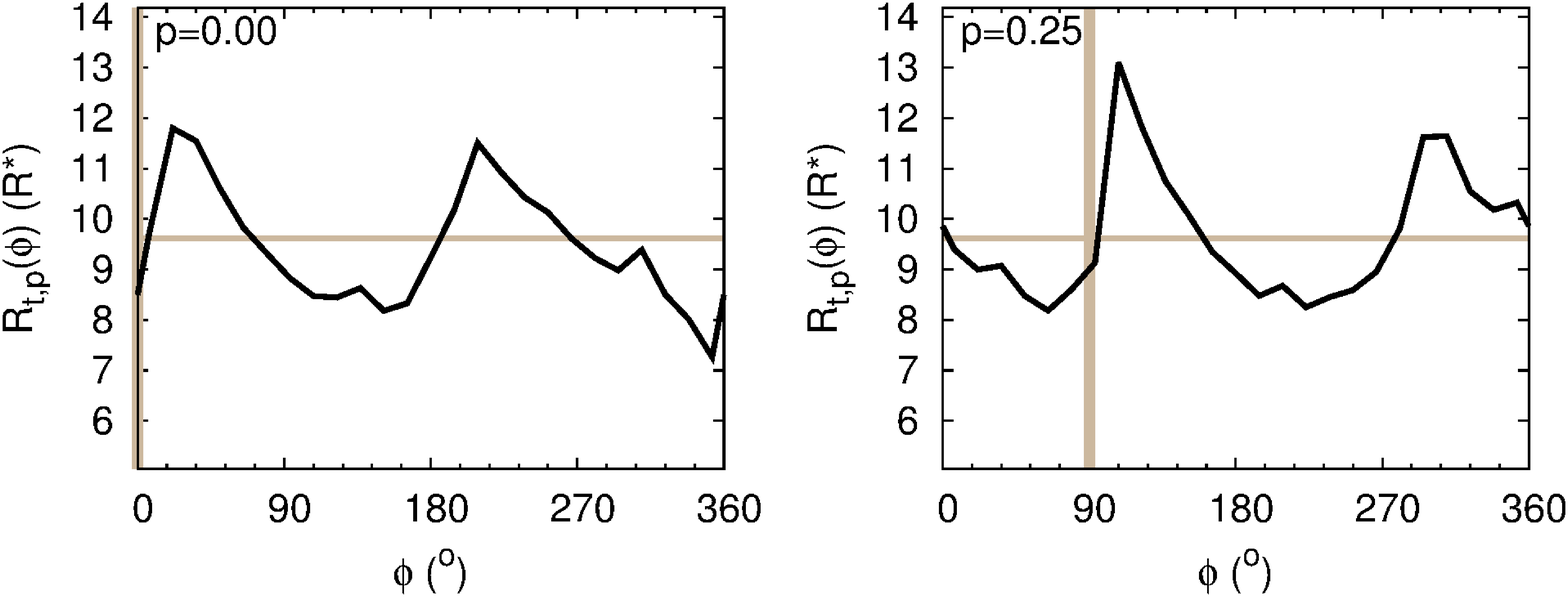}
\caption{The truncation radius $R_{t,p}(\phi)$ as a function of the azimuthal angle $\phi\in[0,2\pi)$ at two different phases. The horizontal line is the mean value and the vertical line shows the direction of the secondary.}
\label{f:rt}
\interspace
\includegraphics[scale=.17]{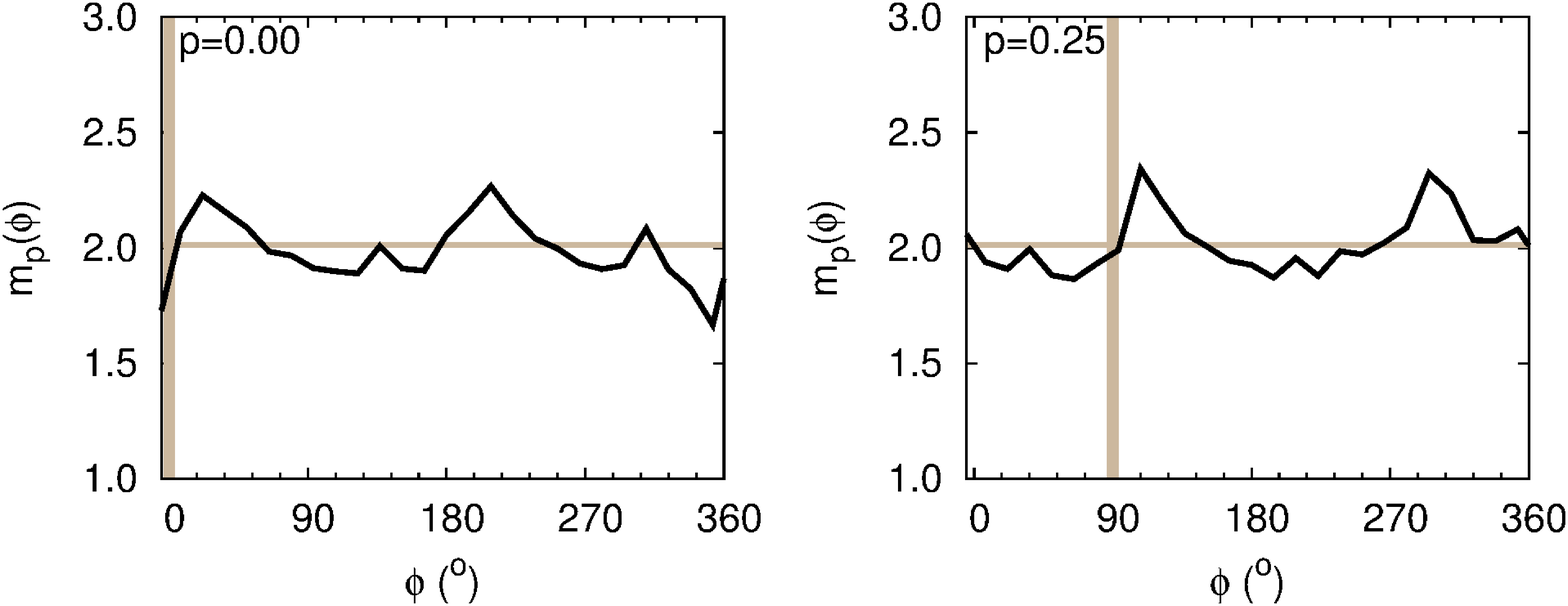}
\caption{Same as Fig.~\ref{f:rt} for the exponent $m_p(\phi)$ of the inner disc.}
\label{f:Mm}
\end{figure}

\begin{figure}\centering
\includegraphics[scale=.17]{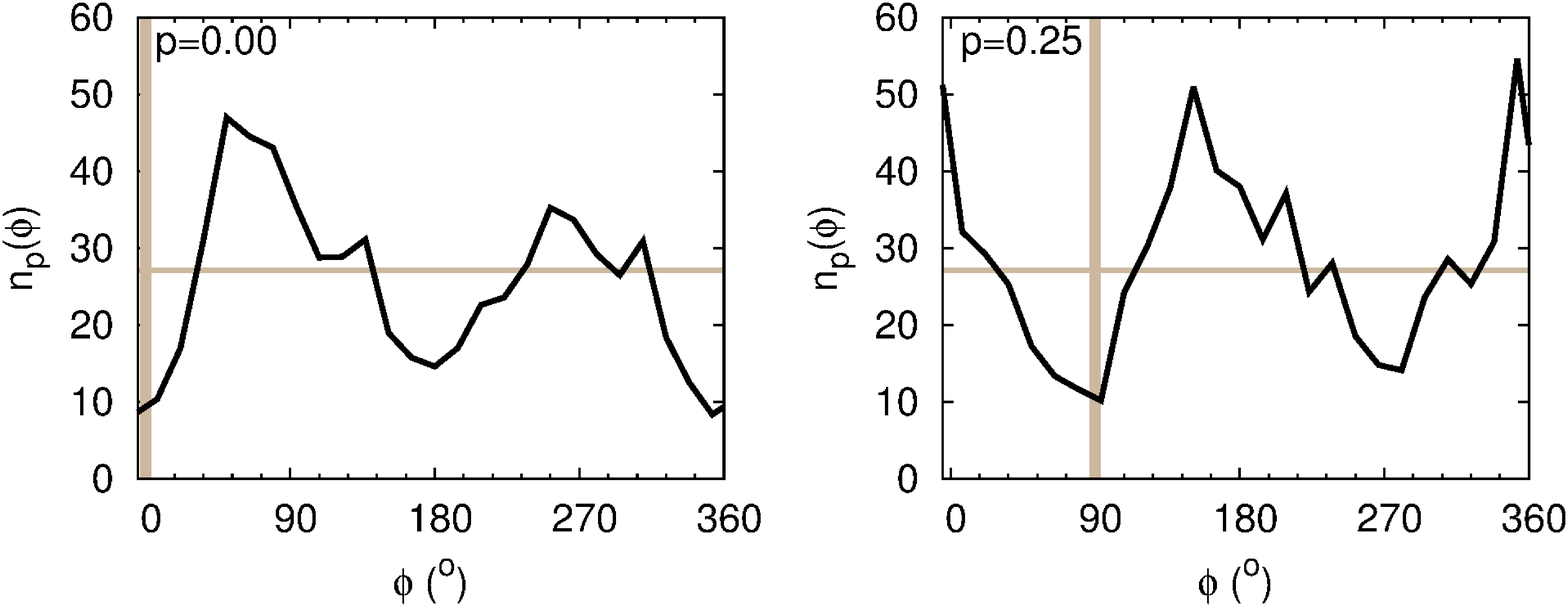}
\caption{Same as Fig.~\ref{f:rt} for the outer exponent $n_p(\phi)$.} \label{f:Nn}
\interspace
\includegraphics[scale=.17]{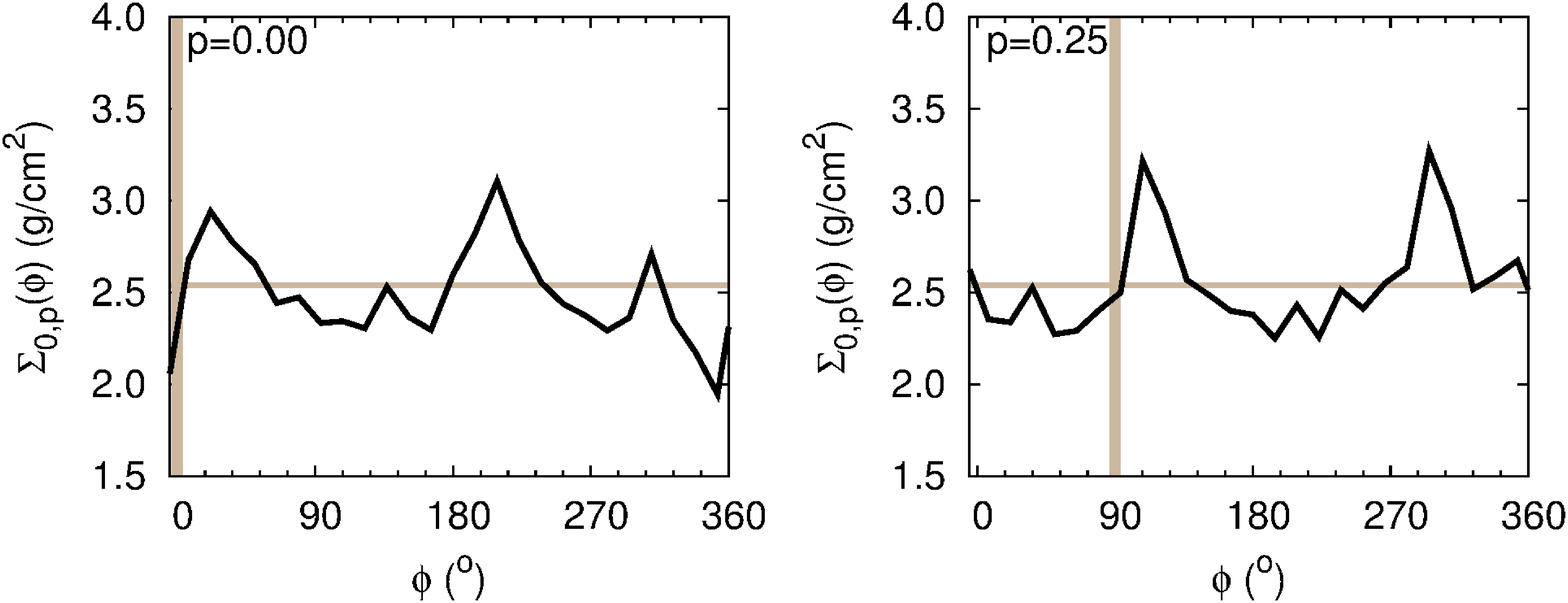}
\caption{Same as Fig.~\ref{f:rt} for the density scale $\Sonep$ at the base of the disc.}
\label{f:Afunc}
\end{figure}

The results of the fitting procedure are shown in figures~\mbox{\ref{f:41-16}-\ref{f:Afunc}}.
Figure \ref{f:41-16} contains polar plots of the truncation radius for the first two snapshots of Fig.~\ref{f:aaec1}.
Figures \ref{f:rt}, \ref{f:Mm}, \ref{f:Nn} and \ref{f:Afunc} plot, respectively, the fitted values for the truncation radius $\Rtpp$, the inner, $m_p(\phi)$, and outer, $n_p(\phi)$, power-law exponents, and the surface density $\Sonep$ at the base of the disc (given by Eq.~\ref{e:Atsne} evaluated at $r=R_*$), as functions of $\phi$ for each phase.

From Fig.~\ref{f:aaec1} we can see that in circular binaries a constant structure is maintained in the disc. Along the orbital cycle it is not modified but only rotates in phase with the position of the companion. Figures \ref{f:rt}-\ref{f:Afunc} prove that for a difference in phase of $\Delta p$, the azimuthal structure is simply shifted by the corresponding amount, $\Delta\phi=\Delta p\cdot360\degr$. The disc attains an orbital phase-locked configuration.

In our simulations no density wave is excited, so that the disc structure is locked with respect to the secondary once the quasi-steady state is reached. From Fig.~\ref{f:41-16}, we notice an elongated structure oriented about $20\degr$ ahead of the secondary. From Fig.~\ref{f:rt}, we see a clear dependence of the truncation radius on $\phi$, with the highest values preceding the angular position of the secondary by about $20\degr$ and $200\degr$, regardless of the phase.
This is further evidence for phase-locking of the surface density. Such a phenomenon is known for decades as the tidal deformation of a Be disc \citep[also seen in][]{OkBa02}: The dominant component of the tidal potential is the 2nd Fourier component that gives this bi-symmetrical structure.

At any phase, the azimuthal direction $\phiel$ of highest elongation (Fig.~\ref{f:rt}) coincides with the angle where the base density also reaches its maximum (Fig.~\ref{f:Afunc}). Moving radially outwards at $\phi_\text{el}$, right out of the surface of the Be star the base density drops more steeply. Hence the fall-off exponent is also higher until the truncation radius (in absolute values) (Fig.~\ref{f:Mm}).
Outward the spiral arm the surface density drops less steeply, as the bulk of the mass corresponding to this azimuthal direction has already been consumed in the dense inner disc and in the more pronounced spiral arm. Since $m_p(\phiel)$ is higher, the surface density is already very low at $r=\Rtp(\phiel)$.
The maximum of the outer disc exponent comes after the direction of elongation (at $\phi>\phiel$, see Fig.~\ref{f:Nn}), where the truncation radius is so close to the star that at this radial distance the density is still very high.

We stress the small azimuthal dependence of the inner exponent $m_p(\phi)$, but it is very small (not visible in the plots of Fig.~\ref{f:cont41}) as compared to the $\phi$-dependence of the outer disc, because the tidal force is much weaker in the inner than in the outer region.

\section{Prograde circular orbits}\label{s:procrc}
In order to investigate the parameter space in terms of selected binary (orbital period, mass ratio, eccentricity) and disc (kinematic viscosity) parameters, simulations with different values of those parameters will be systematically compared to the reference case.
In this section we present our simulations for prograde circular orbits \mbox{($\epsilon=0$)}.

In most Be binaries the periods are longer than a month.
From the representative values of binary parameters presented in \cite{Harm88}, as well as the online catalogue for Be/X-ray binaries and candidates given by \cite{RaPo05}\footnote{\href{http://xray.sai.msu.ru/~raguzova/BeXcat/}{http://xray.sai.msu.ru/$\sim$raguzova/BeXcat/}} for binaries where the companion is a compact object (neutron star or white dwarf), we see that the shortest period ever reported is 12 days.
But this is a highly uncommon case, as there are only four stars with periods between 10-20~days, and six with periods between 20-30 days; all other 80 Be/X-ray binaries and candidates have longer periods.
Excluding such extreme cases, 30~days is a reasonable lower limit for orbital periods of \mbox{Be/X-ray} binaries. Note also that the shortest known period for non-compact companions is 28 days for the Be/sdO binary stars $o$~Puppis \citep{Koub12} and 59~Cygni \citep{PPGT13}.

The range we adopted for the orbital period is between \mbox{5-133} days. This corresponds to the semi-major axis lying in the range between \mbox{28-252~$R_\odot$} (5-46~$R_*$) for the given total mass of the binary system of $\zeta$~Tauri ($M_1+M_2=12.14M_\odot$). The low values of periods (\mbox{5-30} days) were explored only partially, in order to investigate whether there is a minimum separation for which a disc can be present.
We also varied the viscosity parameter within the range \mbox{0.1-1.0}, and the binary mass ratio between 0.08-1.00.

\subsection{Azimuthally averaged properties}
\subsubsection{Dependence on the viscosity parameter}\label{s:deepass}

Figure \ref{f:tempo}b shows the temporal evolution of the azimuthally averaged surface density of a simulation similar to the reference case, but with a much higher viscosity parameter, $\ass=1$. A quasi-steady state is reached much sooner, in already about 21 orbital cycles (half of the time until the steady state of the reference simulation with $\ass=0.4$).
As explained in Section \ref{s:teev}, the time needed to fill up and stabilise the disc depends on $\ass$; large values of the viscosity parameter allow for a much shorter outflow (or diffusion) time-scale.
Another apparent feature of Fig.~\ref{f:tempo}b is that the high-viscosity simulation is truncated further away from the star because of the larger outflow speeds.

\begin{figure}\centering
\includegraphics[clip,trim=0cm 1cm 0cm 1cm,scale=.17]{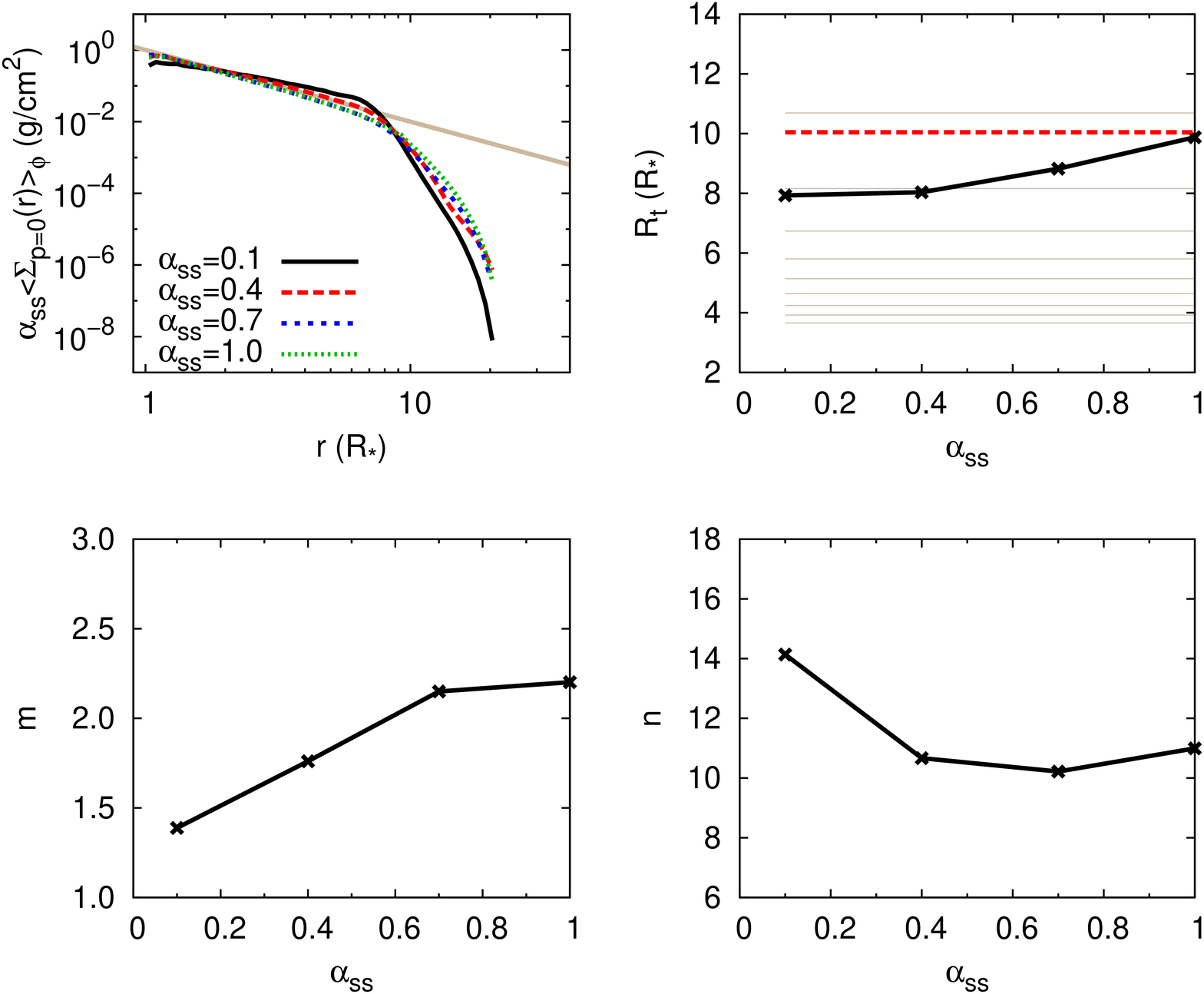}
\caption{The density structure at quasi-steady state for our reference system ($\ass=0.4$) and three other values of $\ass$ (0.1, 0.7, 1.0).
\emph{Top left:} the product of the viscosity parameter and the azimuthally averaged radial surface density, $\ass\Sradz$, at $p=0$. The straight line shows the density profile of the decretion disc of an isolated Be star (Eq.~\ref{e:sd}).
\emph{Top right:} The truncation radius fitted with Eq.~\eqref{e:Atsuo} (solid line). The thin straight lines denote the 2:1, $\ldots$, 10:1 resonance radii (from top to bottom, respectively) and the Roche radius from Eq.~\eqref{e:Egg} is plotted as the dashed (red) line.
\emph{Bottom left:} power-law exponent for the inner part of the disc. \emph{Bottom right:} outer power-law exponent.}
\label{f:ass}
\end{figure}

The top left panel of Fig.~\ref{f:ass} shows the surface density structure of the disc for the reference simulation and three other simulations with different values of $\ass$. The remaining panels show the fitted values (with Eq.~\ref{e:Atsuo}) for $R_t$, $m$ and $n$, as functions of $\ass$ of each simulation.
The truncation radius increases with $\ass$, and is around the 3:1 resonance ratio (where $r/a=0.48$), a result that agrees with the conclusions of \cite{OkNe01b} for low eccentricities.
We also notice that the Roche lobe radius of the primary Be star as given by the Eggleton formula lies between the 2:1 and the 3:1 resonance radii and is always a bit larger than the truncation radius.

In order to distinguish the similarities and differences of the decretion disc structure in a binary system compared to that of an isolated star, apart from the existence of truncation, we compare the power-law exponent of the disc inward the truncation.
For $\ass\ge0.6$ the inner disc has a slope close to that of a disc around an isolated Be star ($m\gtrsim2$), but for lower viscosity values the inner slope is much lower ($m=1.2$ for $\ass=0.1$).
This density enhancement in the inner disc was also noticed by \cite{OkBa02}, who attributed it to the resonant torque stopping the angular momentum from being transported outwards, thus causing the accumulation of material in the inner region.
Henceforth we call this phenomenon the \emph{accumulation effect}.

The outer disc exponent $n$, which quantifies the steepness of the outer disc, seems to decrease with increasing values of viscosity, i.e.~the surface density outward the truncation radius becomes shallower. However this is of minor significance, because the outer disc structure keeps varying a little with time, even when the inner region of the disc has reached a quasi-steady state.
Moreover, the density in the outer disc is very low, and these timely variations do not modify e.g.~the total disc mass at steady state.
Also, as discussed in Section~\ref{s:pseudo}, the outer disc is unlikely to contribute significantly to the emergent spectrum, thus bearing no effect on the observables.

\subsubsection{Dependence on the orbital period}\label{s:depprb}
Figure \ref{f:tempo}c shows the temporal evolution of a long period (\mbox{$\Porb=133$} days) binary system. We notice that the steady state is reached in about 24 periods (3\,200 days or $\sim8.7$~years), about three times longer than the time to steady state of our reference simulation of $\Porb=30$ days. The longer time to reach the steady state is a result of the larger truncation radius.

\begin{figure}\centering
\includegraphics[clip,trim=0cm 1cm 0cm 0cm,scale=.17]{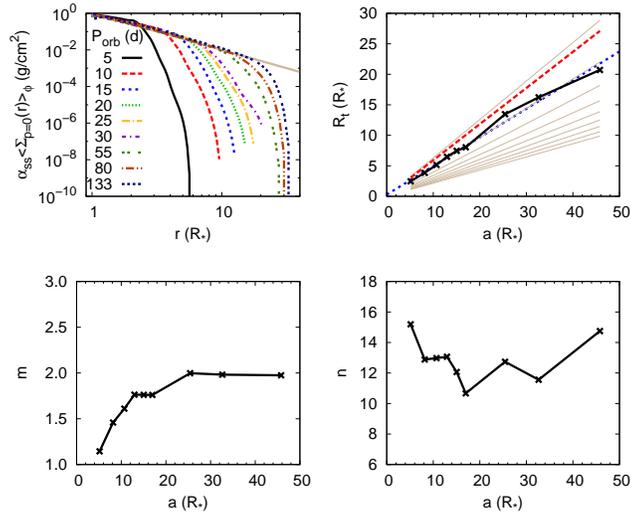}
\caption{Same as Fig.~\ref{f:ass}, but for systems with different orbital periods in the range 5-133 days ($a\in[5.1,45.8]R_*$). The dotted (blue) line is a linear fit of $R_t$ vs.~$a$.}
\label{f:porball}
\end{figure}

Figure \ref{f:porball} shows how the surface density of the disc changes with varying orbital periods/separations. The truncation radius grows linearly with the orbital separation, and almost coincides with the 3:1 resonance radius.
This again confirms the conclusion of \cite{OkNe01b} with respect to low-eccentricity orbits.
The resulting truncation radius is always much smaller than the orbital separation, as found in \cite{NeOk01a} and \cite{OkNe01a}, while the positive correlation between the two confirms the relevant conclusions of \cite{ReFC97}.
We find that $R_t\approx0.8\RLBe$, which suggests that Eq.~\eqref{e:Egg} (here evaluated for the Roche lobe radius of the Be star $\RLBe$, reversing the mass ratio) is a good first order approximation for $R_t$.

The accumulation effect is very much dependent on the orbital separation. It is quite strong for short-period binaries, disappears for binary separations $a\gtrsim 25R_*$ (\mbox{$\Porb\gtrsim55$~days}), and remains largely constant thereafter.

A question that one might ask is what the lowest orbital period is, for which a disc is still present. 
In other words, would the absence of low-period Be binaries be a selection effect caused by disruption of the disc due to a close companion?
Our results indicate that a disc, albeit small, is present even for very close binaries. For instance, the system with \mbox{$\Porb=5$~days} still has a disc with $R_t=2.5R_*$, which should be easily observable. An extrapolation of the function of $R_t(a)$ to even lower values of $a$ shows that $R_t$ becomes equal to $R_*$, i.e.~the disc vanishes, when $a\simeq R_*$.
Therefore, we conclude that the absence of Be stars with short periods is a real phenomenon, and not a result of the absence of discs in low-period binaries.

\subsubsection{Dependence on the binary mass ratio}

Figure \ref{f:tempo}d shows that circular binaries with a high mass ratio need longer time to reach the quasi-steady state (about 55 orbital cycles, i.e.~4.5 years). That is because the gravitional forces in this case are much stronger, many more particles remain bound to the system (with the actual decretion rate smaller than all other simulations; see also Table \ref{t:chars}), resulting to a more massive disc that needs more time to relax.

\begin{figure}\centering
\includegraphics[scale=.17]{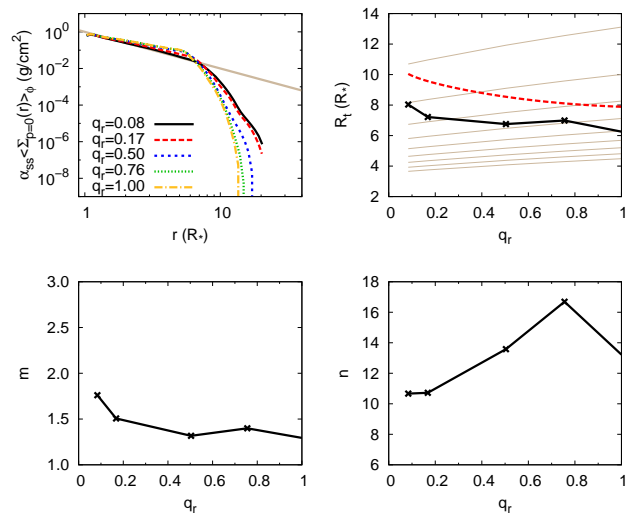}
\caption{Same as Fig.~\ref{f:ass}, but for simulations of the reference case and of other binary mass ratios (0.08, 0.17, 0.50, 0.76, 1.00).}
\label{f:qmass}
\end{figure}

Figure \ref{f:qmass} shows the results for simulations of different mass ratios; it is clear that the larger the mass ratio, the larger the accumulation effect. The accumulation effect is always present, since both the value of viscosity ($\ass=0.4$) and of the orbital separation ($a=17R_*$, \mbox{$\Porb=30$}~days) lie within the region where the accumulation effect is exhibited ($\ass<0.6$, $a<25R_*$, $\Porb<55$~days) for our chosen values of stellar parameters.
We also confirmed \cite{WhKi91} in that the resonance 3:1 can appear only for mass ratios $q_r\le0.25$, while for larger mass ratios the available resonances are weaker. As before, we find that $R_t\approx0.8\RLBe$ is a good approximation for the truncation radius.

\subsection{Phase dependence and azimuthal structure}
Similarly to the procedure followed in Section \ref{s:phase_dependent}, we analysed the azimuthal structure of the circular binary systems examined in the preceeding section. We find that, irrespectively of the value of the parameters explored ($\Porb$, $q_r$, $\ass$), the same conclusions hold:
The surface density attains a phase-locked behaviour, in which the {\twod} structure rotates in phase with the secondary, the disc becomes elongated with the maximum elongation preceding the secondary, while the inner inner exponent $m_p(\phi)$ shows small variation with $\phi$.

\begin{figure}\centering
\includegraphics[scale=.17]{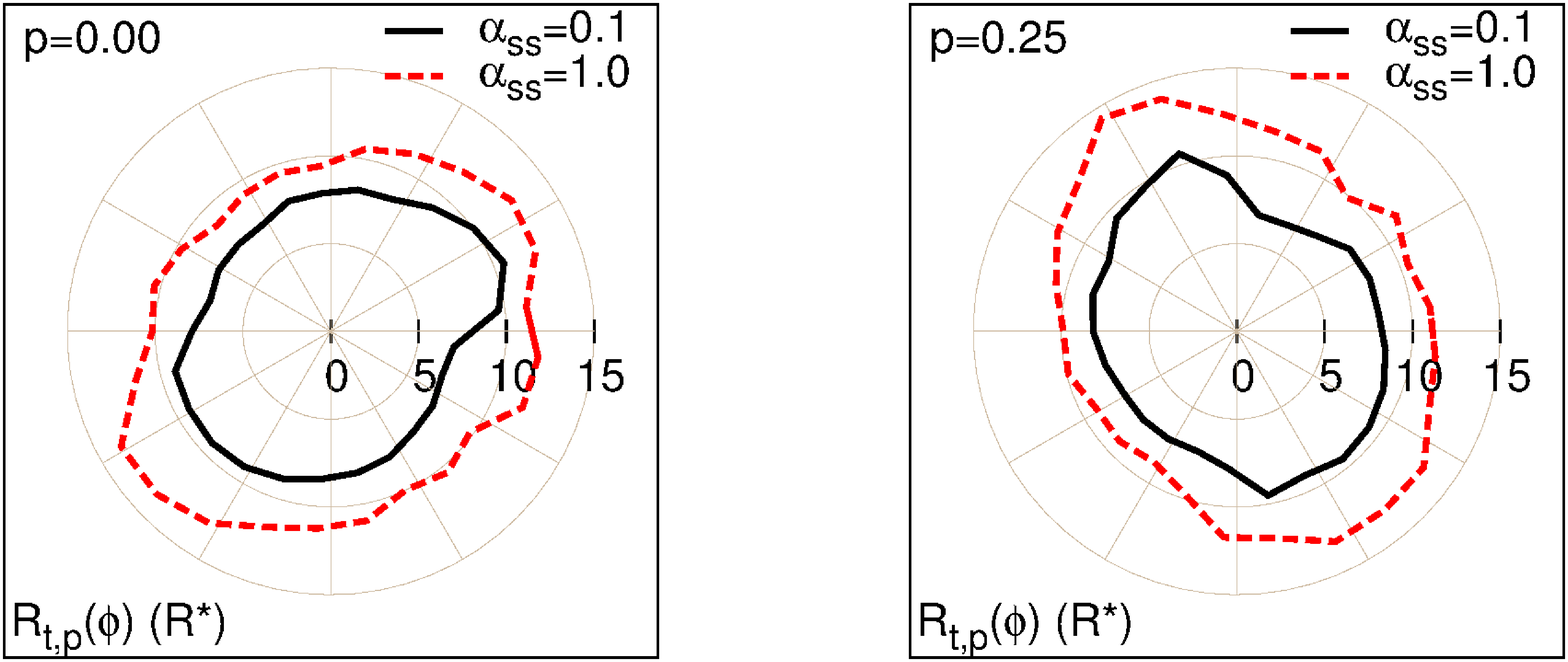}
\caption{Same as Fig.~\ref{f:41-16}, but for two simulations with different values of $\ass$ (0.1, 1.0), with all other parameters equal to the ones of the reference system.}
\label{f:alphas}
\interspace
\includegraphics[scale=.17]{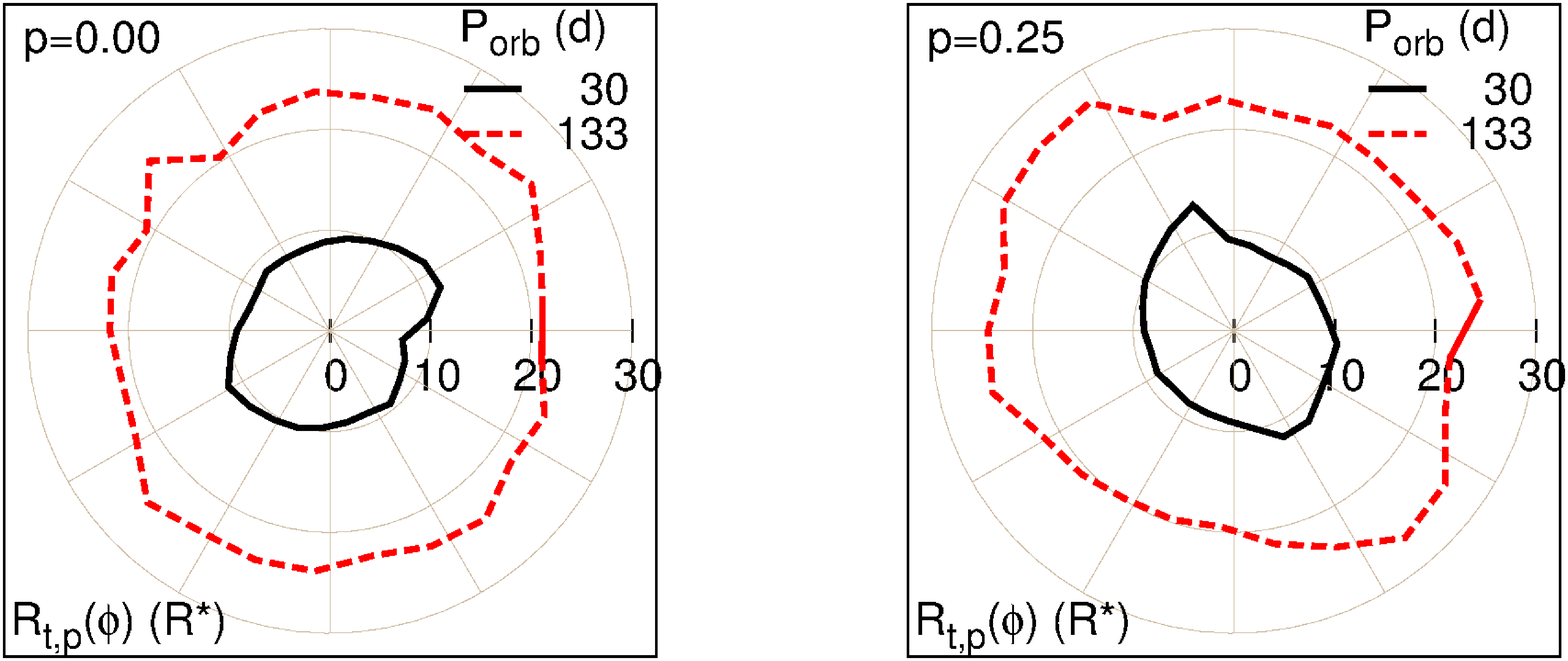}
\caption{Same as Fig.~\ref{f:41-16}, but for two simulations with different values of $\Porb$ (30 and 133 days), with all other parameters equal to the ones of the reference system.}
\label{f:porbs}
\interspace
\includegraphics[scale=.17]{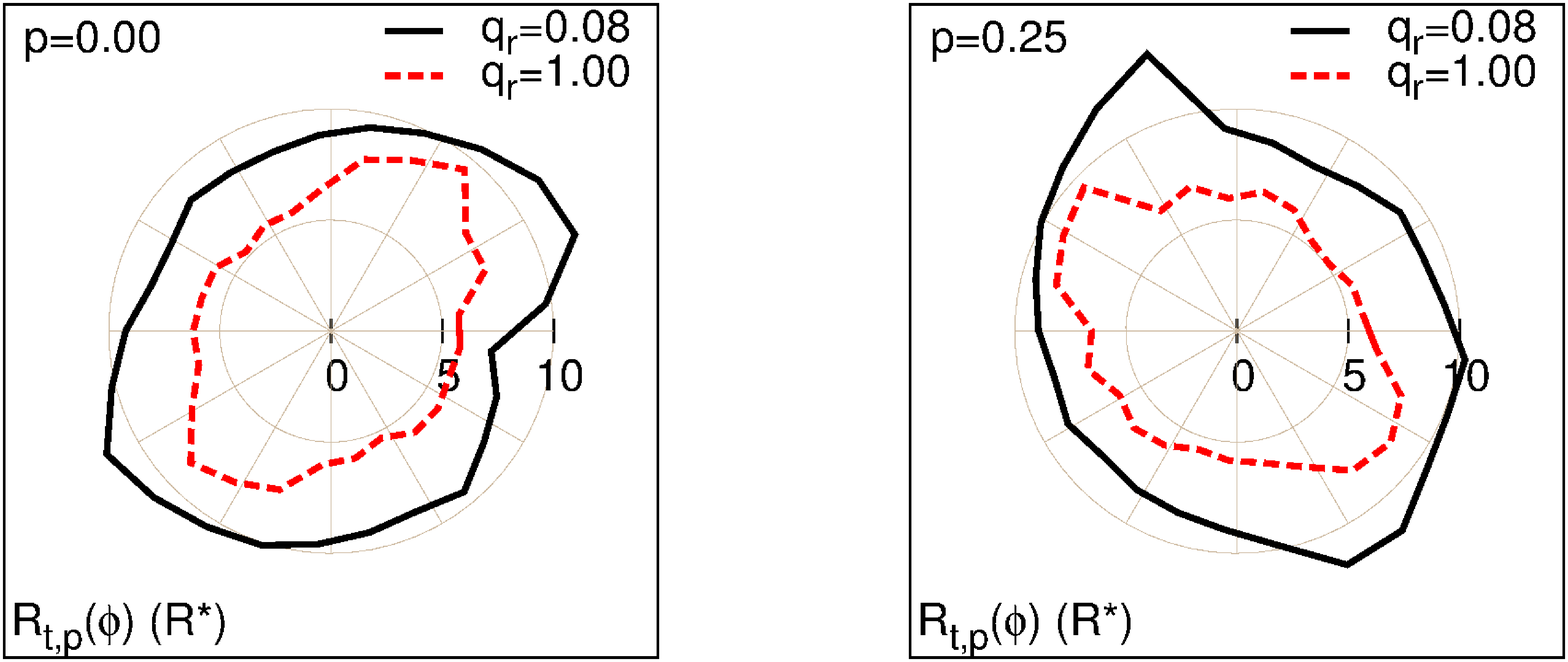}
\caption{Same as Fig.~\ref{f:41-16}, but for two simulations with different values of $q_r$ (0.08, 1.00), with all other parameters equal to the ones of the reference system.}
\label{f:qrs}
\end{figure}

A higher value of the viscosity parameter and/or of the orbital period results in a more axisymmetric structure (the spiral arms are not so pronounced). This generally agrees with \cite{OkBa02} in that the orbital modulation minimises for higher values of $\alpha$.
The contrary happens with smaller viscosity parameter (Fig.~\ref{f:alphas}), smaller period (Fig.~\ref{f:porbs}) and/or higher secondary-to-primary mass ratio (Fig.~\ref{f:qrs}): smaller disc extent, more notably elongated structure, and lower decretion rates (Table \ref{t:chars}).

As shown in Fig.~\ref{f:alphas}, the time lag appears to increase with increasing values of $\ass$. For $\ass=1$ we find that the time lag is larger than the simulation with $\ass=0.4$, and is approximately equal to $30\degr$; for $\ass=0.1$ the time lag with respect to the position of the secondary is smaller, \mbox{$\simeq10\degr$}.
The time lag for a higher mass ratio ($q_r=1$, Fig.~\ref{f:qrs}) is larger, $45\degr$. About the same value we estimated for the time lag of maximum elongation with respect to the secondary for the simulation with $\Porb=133$ days, but the disc is so much axisymmetric in this case that it is meaningless to talk about a time lag.

\section{Prograde elliptical orbits}\label{s:proell}
In this section we explore the surface density structure for systems of elliptical orbits ($0<\epsilon \le 0.9$). These configurations may arise either when the binary system has been formed by the capture of one star by another or during an asymmetric supernova explosion that changes the orbits from circular to elliptical.
Such events occur randomly and therefore the orbits might not only be highly eccentric, but also retrograde and misaligned (meaning that the inclination angle between the spin axis of the central star and the orbital axis is not zero; \citealt{Lai97}).

\begin{figure}\centering
\includegraphics[width=\linewidth]{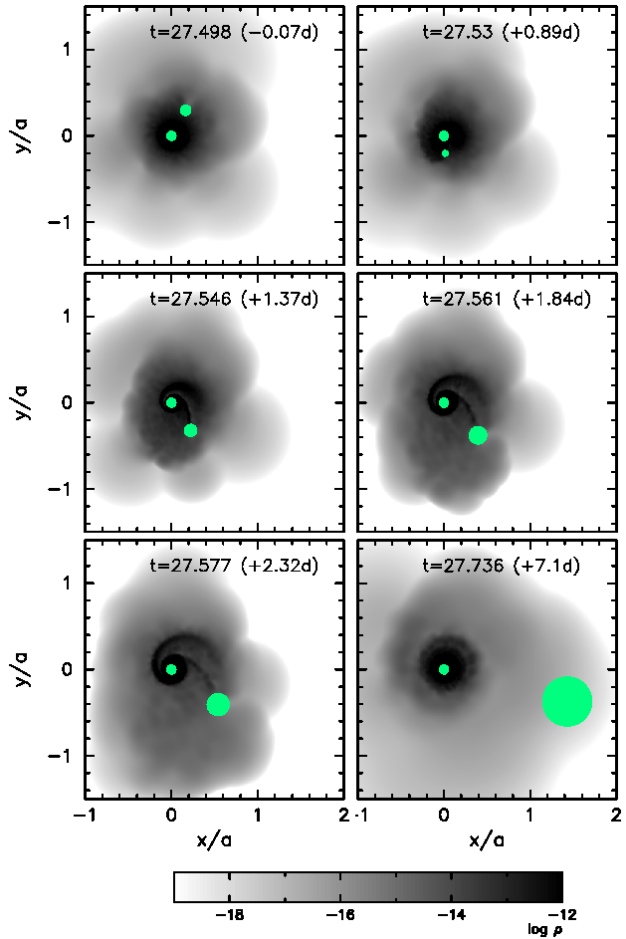}
\caption{Same as Fig.~\ref{f:aaec1}, but for a simulation of an eccentric binary system ($\epsilon=0.9$) at six different phases.}
\label{f:aaec2}
\end{figure}

Unlike the, constant in shape, rotating structure of the disc in circular systems, we expect that eccentric orbits stir up changes in the azimuthal structure of the disc with the orbital phase.
A higher eccentricity causes the influence of the secondary on the disc to vary with the orbital phase, and the two stars interact effectively for a smaller time within each orbit.
Figure \ref{f:aaec2} shows snapshots at six different orbital phases of a system similar to our reference system, but for $\epsilon=0.9$. Note that the size of the secondary (represented by its accretion radius) now changes along the orbital cycle, as the distance $D$ between the two stars is now variable (Eq.~\ref{e:Egg}).

The simulation evolved long enough until no relevant cycle-to-cycle variations were present.
The time to steady state was in general shorter than other cases (Fig.~\ref{f:tempo}e). Figure~\ref{f:Mdisc} shows that increasing eccentricity results in lower total disc mass, whose asymptotic value is reached sooner. Particularly for the simulation with the highest eccentricity, the disc mass oscillates with period equal to the orbital phase (Fig.~\ref{f:Mdisc}).
At every cycle the disc is heavily disrupted after the periastron, and the remaining time until the next periastron the disc is being rebuilt. This remaining time is not enough to reach the mass of the circular binary, and this results in very low densities and few particles (see also Table~\ref{t:chars}).
The snapshots do show a disc, but the total disc mass is about one order of magnitude lower than the disc mass in equivalent circular systems.

\begin{figure}\centering
\includegraphics[clip,trim=0mm 1cm 0mm 1cm,scale=.17]{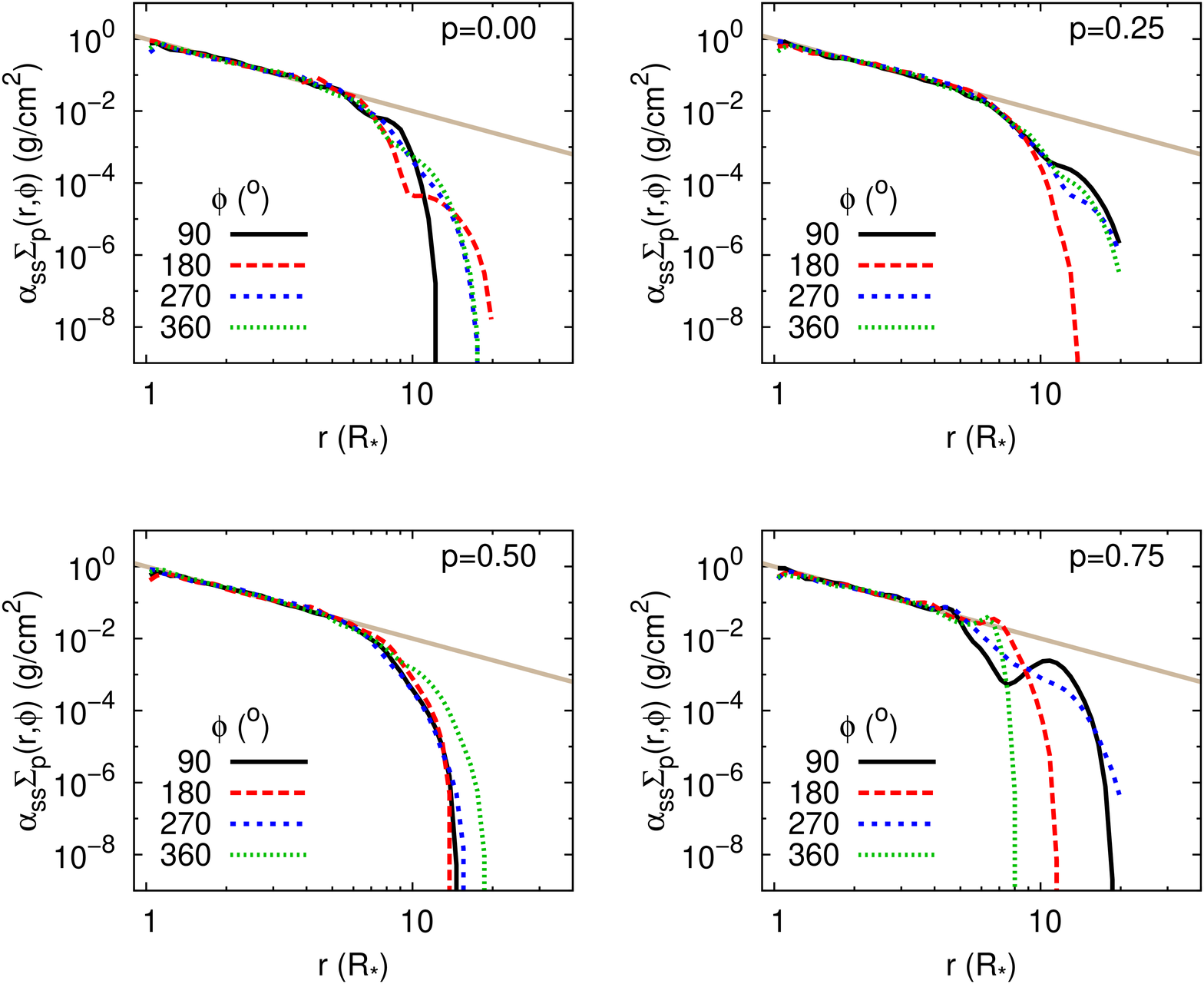}
\caption{Same as Fig.~\ref{f:cont41}, for a simulation similar to the reference system but for $\epsilon=0.3$.}
\label{f:e03}
\interspace
\includegraphics[clip,trim=0mm 1cm 0mm 1cm,scale=.17]{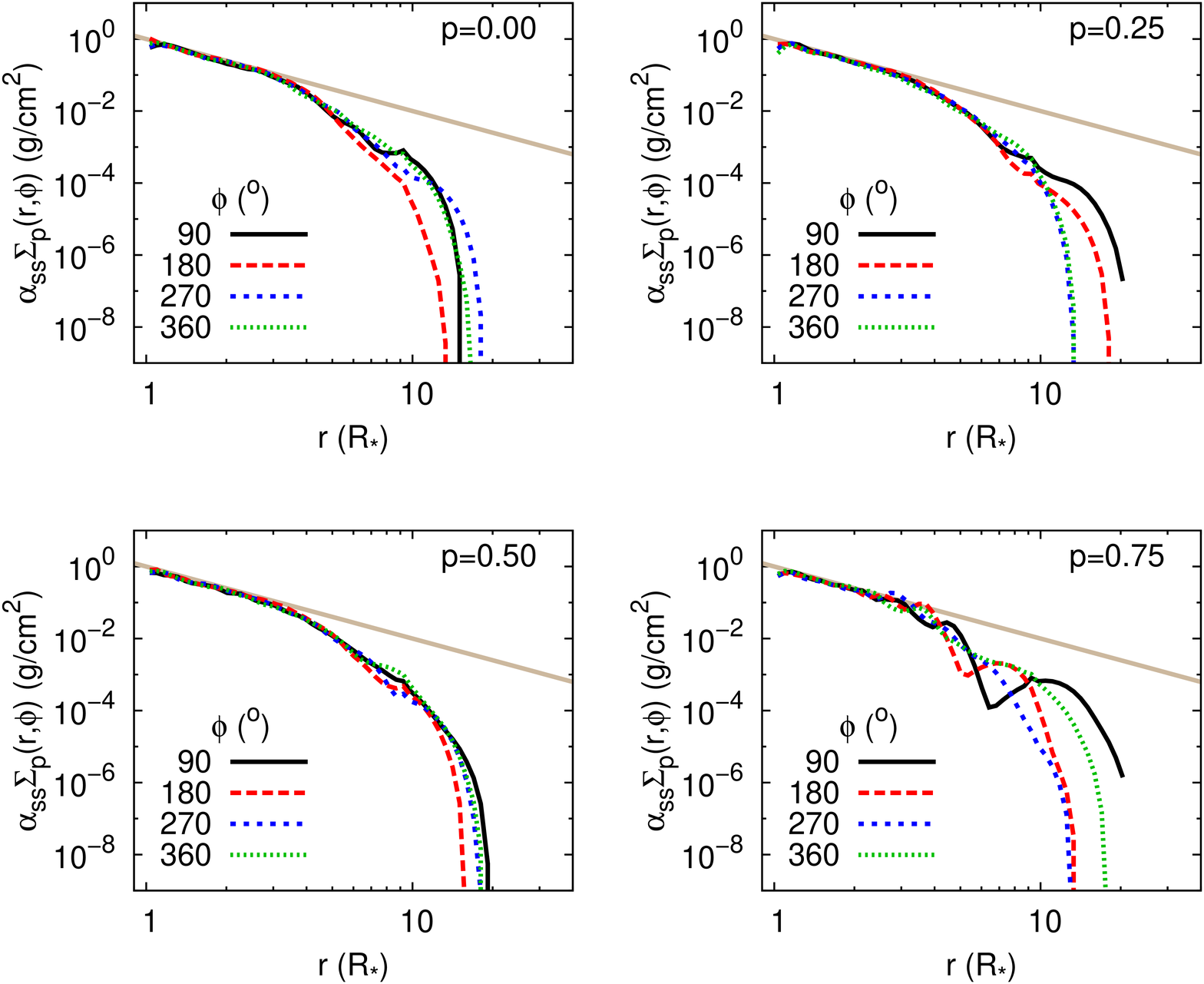}
\caption{Same as Fig.~\ref{f:cont41}, for a simulation similar to the reference system but for $\epsilon=0.6$.}
\label{f:e06}
\end{figure}

\begin{figure}\centering
\includegraphics[clip,trim=0mm 1cm 0mm 9mm,scale=.17]{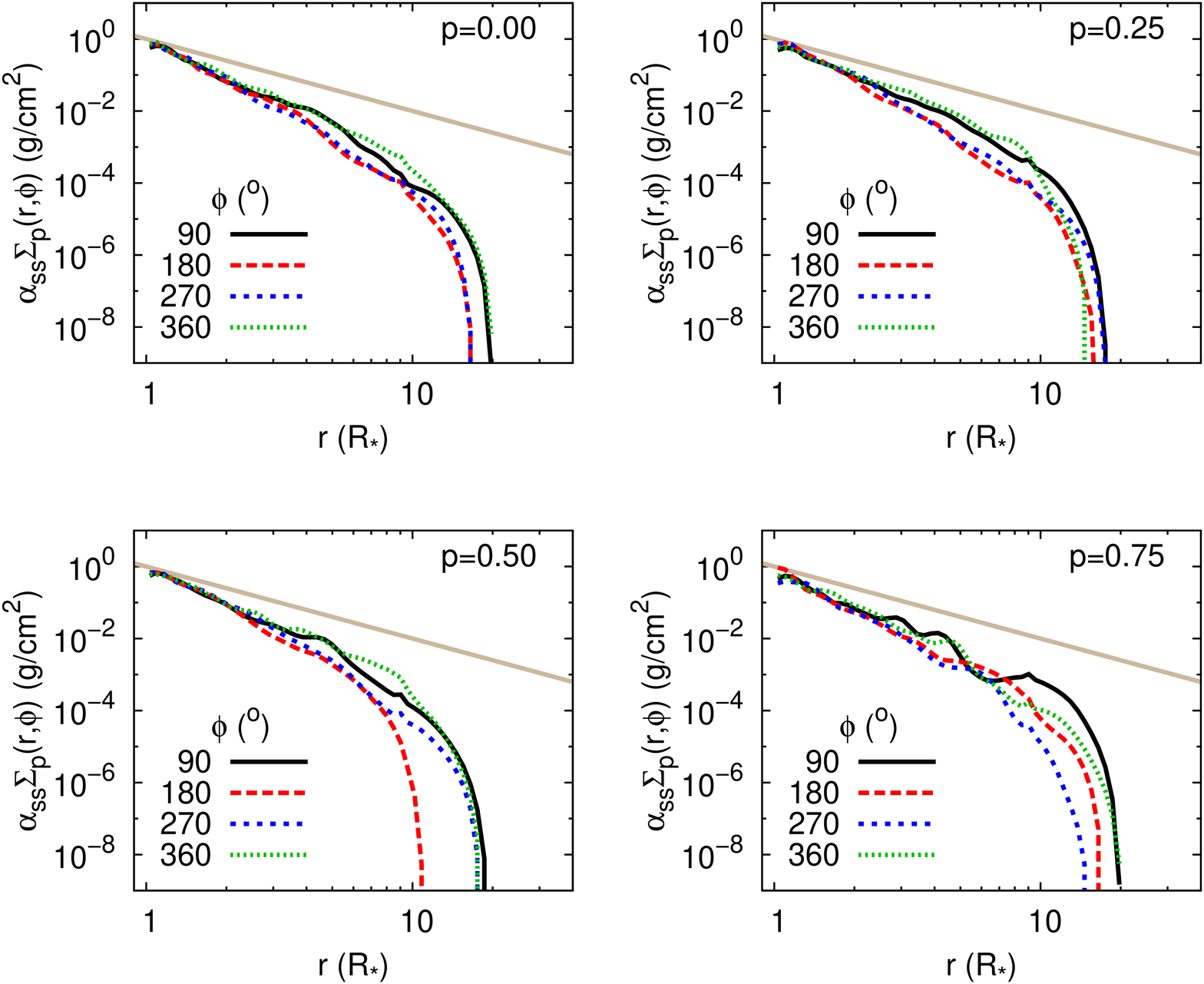}
\caption{Same as Fig.~\ref{f:cont41}, for a simulation similar to the reference system but for $\epsilon=0.9$.}
\label{f:e09}
\end{figure}

At apastron and even a little after the periastron the disc is nearly axisymmetric. Its structure is strongly perturbed after the periastron, where the disc develops a marked two-armed density wave; this characteristic has also been reported by \cite{OkBa02}.
Around phase $p\simeq0.8$ in this particular simulation, the disc becomes more circular due both to phase mixing and the effect of viscosity. In lower eccentricities the two spiral arms persist for a longer period along each orbital cycle, and this phenomenon degrades to the arms being constant in the circular case, as we saw in Section \ref{s:phase_dependent} (Fig.~\ref{f:aaec1}).

The {\twod} disc structure is investigated in detail in figures \ref{f:e03}, \ref{f:e06} and \ref{f:e09}, for three different eccentricities ($\epsilon = 0.3$, 0.6 and 0.9, respectively).
Similarly to Fig.~\ref{f:cont41}, the surface density profile is plotted for four values of $\phi$ and four orbital phases $p$.
In the outer parts the disc displays a complex behaviour, changing both azimuthally and as the secondary follows its orbital path. The strongest perturbations occur after the periastron passage, when the spiral arms are formed.
However, the most important feature in these figures, because of its observational consequences (see Section \ref{s:pseudo}), lies in the effect of the eccentricity in the inner disc region.

The low-eccentricity system (Fig.~\ref{f:e03}) is similar to the (circular) reference system: the density of the inner part of the disc ($r\lesssim6R_*$) follows a power-law whose slope is largely independent of $\phi$, i.e.~the inner region remains axisymmetric, regardless of the orbital phase. As eccentricity increases, the slope of the inner disc region increases (especially for $\epsilon>0.6$), and is now more dependent on $\phi$, oscillating over a larger range of values.
The latter is a consequence of the spiral arms, which become more evident in eccentric cases, although they do not survive for long. The truncation radius (defined as the region where the disc structure becomes steeper) increases, but is not accompanied by higher mass due to the low densities.

\begin{figure}\centering
\includegraphics[clip,trim=0mm 1cm 0mm 1cm,scale=.17]{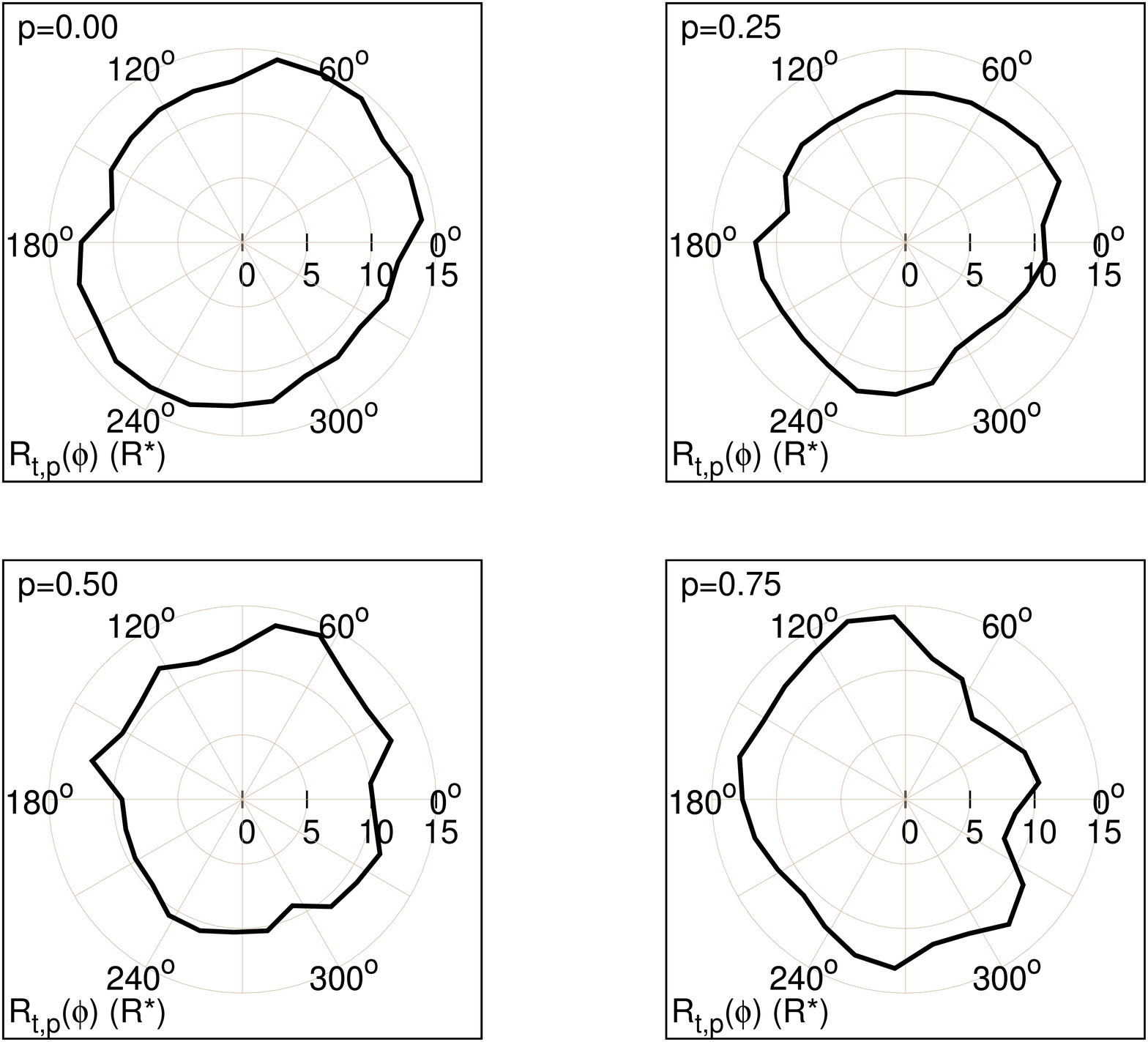}
\caption{Same as Fig.~\ref{f:41-16}, but for a highly eccentric prograde orbit ($\epsilon=0.9$) for four different phases.}
\label{f:91-16}
\interspace
\includegraphics[clip,trim=0mm 1cm 0mm 1cm,scale=.17]{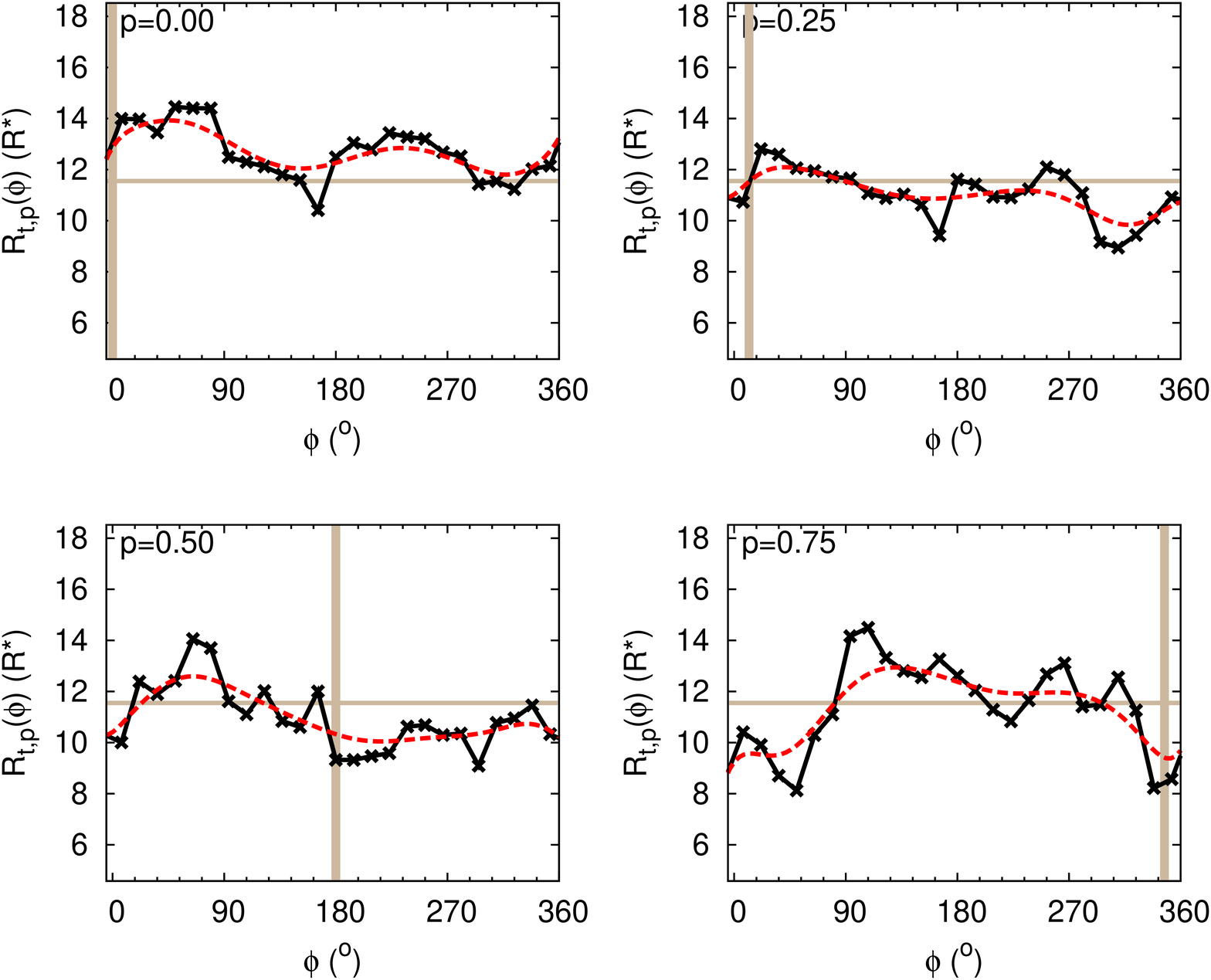}
\caption{Same as Fig.~\ref{f:rt}, but for a highly eccentric prograde orbit ($\epsilon=0.9$) for the four phases depicted in Fig.~\ref{f:91-16}. The dashed (red) line simply smooths the data, in order to show the trends.}
\label{f:91-06}\label{f:rt-vs-phi}
\end{figure}

\begin{figure}\centering
\includegraphics[clip,trim=0mm 1cm 0mm 0cm,scale=.17]{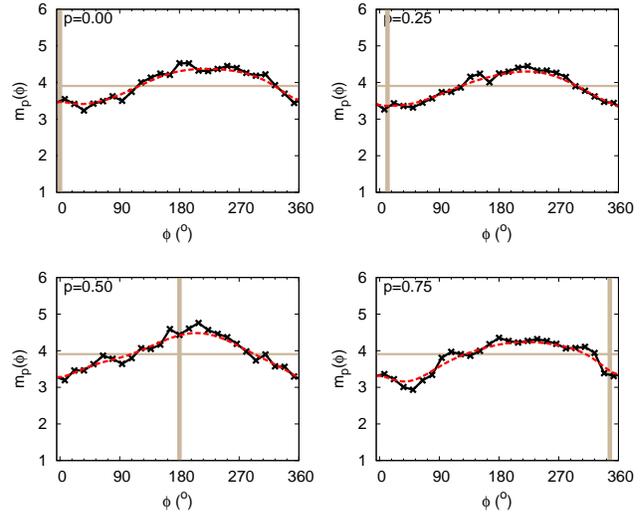}
\caption{Same as Fig.~\ref{f:91-06} for the inner disc exponent.}
\label{f:91-08}
\end{figure}

The above are quantitatively shown in figures~\mbox{\ref{f:91-16}-\ref{f:91-08}}. In particular, figures \ref{f:91-16} and \ref{f:91-06} suggest that in highly eccentric orbits the truncation radius reaches its maximum value at apastron ($p=0.0$) and its minimum at periastron ($p=0.5$). Figure \ref{f:91-08} shows that the inner disc density structure becomes steeper at azimuthal angle $\phi\in(180,270)\degr$, independently of the orbital phase.

Concluding, we define two limiting cases for the disc properties. The discs around slightly eccentric binaries (\mbox{$0\le\epsilon\lesssim0.6$}) have an inner region that is mostly axisymmetric, and whose overall structure does not vary substantially with the orbital phase. 
For highly eccentric orbits the disruption of the disc around periastron results in an inner density slope much steeper than the nearly circular cases (i.e.~$m\gg2$), and also very asymmetric.

\begin{figure}\centering
\includegraphics[width=\linewidth]{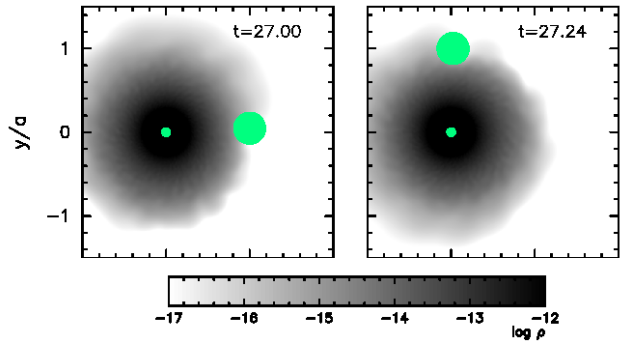}
\caption{Same as Fig.~\ref{f:aaec1}, but for a retrograde and circular binary orbit.}
\label{f:daaac}
\end{figure}

\section{Retrograde orbits}\label{s:rtrcrc}
\subsection{Circular orbits}
We will now explore how binary companions in retrograde orbits affect the disc.
As noted in the previous section, retrograde orbits are usual in the capture scenario for the formation of a binary system (usually highly eccentric orbits).

As discussed in Section \ref{s:model}, truncation occurs when the viscous torque equals the tidal torque.
This usually does not happen in retrograde orbits. That is because the time of interaction between the disc particles and the companion star is short, as they orbit the central star in opposite directions. As a result, a much weaker perturbation of the disc is expected.

This can be verified in Fig.~\ref{f:daaac}, where we present surface density maps of a simulation similar to the reference system, but in retrograde orbital motion. It is clear that the disc is almost circular (in contrast to the elongated shape of the circular prograde system, Fig.~\ref{f:aaec1}), while the disc extends to much larger distances from the star than the prograde case.
Figure \ref{f:tempo}f shows the temporal evolution of the azimuthally averaged surface density.
We don't see any real break in the radial density distribution. Nevertheless, fitting the surface density to Eq.~\eqref{e:Atsuo}, we find that the truncation radius is $R_t=10R_*$, larger than the prograde system.
Observationally, therefore, discs around circular retrograde systems should appear much bigger and show no orbital or azimuthal modulation. The inner exponent is slightly higher, thus the surface density and the total disc mass are a little lower than the prograde system (Table \ref{t:chars}).

\subsection{Elliptical orbits}

The structure of retrograde elliptical systems are different with respect to their prograde counterparts, with a high dependence on both $\phi$ and $p$. The disc is much sparser than all cases analysed previously (see also total disc mass in Table~\ref{t:chars}), and the surface density has dropped below 0.01~g/cm$^2$ already at distances $2-3R_*$.

\begin{figure}\centering
\includegraphics[width=\linewidth]{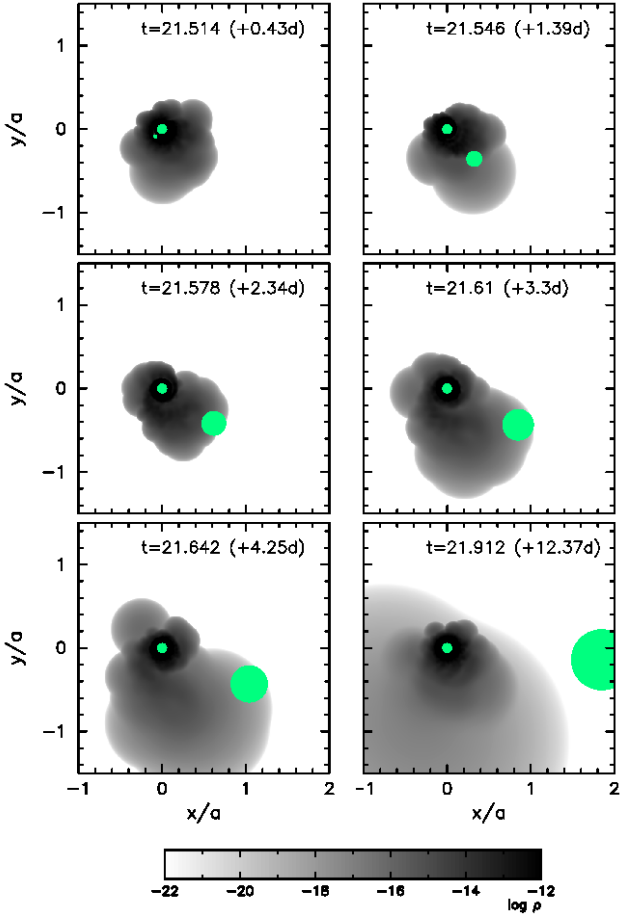}
\caption{Same as Fig.~\ref{f:aaec2}, but for a retrograde and highly elliptical ($\epsilon=0.9$) binary orbit.}
\label{f:dretr}
\end{figure}

\begin{figure}\centering
\includegraphics[clip,trim=0mm 1cm 0mm 1cm,scale=.17]{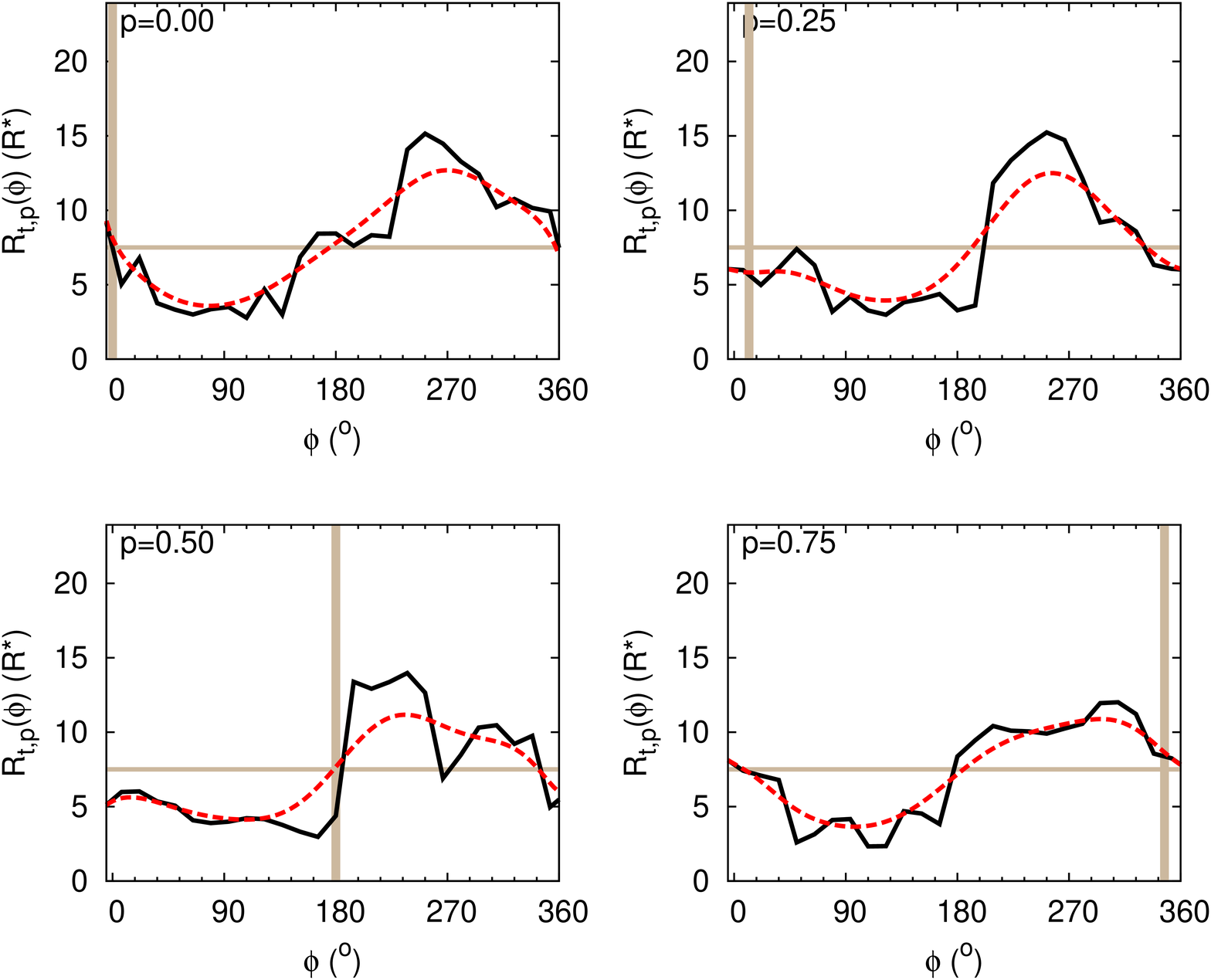}
\caption{Same as Fig.~\ref{f:rt}, for a retrograde and highly eccentric system ($\epsilon=0.9$).}
\label{f:rea}
\end{figure}

From the snapshots of Fig.~\ref{f:dretr}, we see that high eccentricity in retrograde orbits causes very low densities (note the much lower density scale, with respect to the rest of the snapshots presented) and very intense azimuthal dependence of the disc structure, especially after the periastron.
The density structure is one-sided, while its extent maximizes at about the same range of azimuthal angles after the periastron (around $\phi\simeq270\degr$), irrespectively of the orbital phase (see Fig.~\ref{f:rea}). What changes along an orbital cycle is the total disc size, which decreases until the periastron ($p=0.5$), and increases after that, at phase $p=0.6$. At this phase, a small and less dense, with respect to the highly eccentric prograde system (Fig.~\ref{f:aaec2}), density wave is formed, but is destroyed almost immediately.

\section{Disc continuum emission of Be binaries} \label{s:pseudo}
In the previous sections we explored how decretion discs around Be stars are affected by the presence of a binary companion. 
The combination of different orbital and disc parameters causes variations on the disc density slope, on its extent (as measured by the truncation radius) and on its azimuthal structure. In this section we discuss the implications of these modifications on the disc continuum emission.

According to \citet{ViCa15}, the disc can be separated into two components according to their emission properties: an inner optically thick region, called the pseudo-photosphere, and an outer optically thin region. The radius $\Rbar$ of the pseudo-photosphere is defined as the position on the disc where the optical depth $\tauobs$ in the line of sight is close to unity, and is a function of the disc parameters as well as of the wavelength.
For an isothermal disc the specific intensity is uniform inward $\Rbar$, and drops as a power law in the tenuous region.

In most of the simulations discussed previously, the truncation region, defined by the phase- and $\phi$-dependent quantity $\Rtpp$, divides the disc into an inner part \mbox{($r\le R_t$)}, which is much denser and has a less steep density fall-off than the outer part ($r\ge R_t$).
Since $\tauobs\propto\Sigma^2$, the disc emission beyond $R_t$ is likely negligible, therefore continuum observations will only probe the disc inside the truncation region.

For nearly circular orbits, the truncation region does not practically depend on the orbital phase, and the disc keeps its azimuthal symmetry. Apart from truncation, the main effect caused by the binary interaction is the accumulation of material inside the truncation region, causing the inner disc density exponent to become lower than the steady-state value for isolated Be discs.
A less steep density profile moves the position of $\Rbar$ outwards, since $\Rbar<R_t$ and the entire optically thick region becomes denser. 
Therefore the disc becomes brighter with decreasing $m$.

\begin{figure}\centering
\includegraphics[scale=.3]{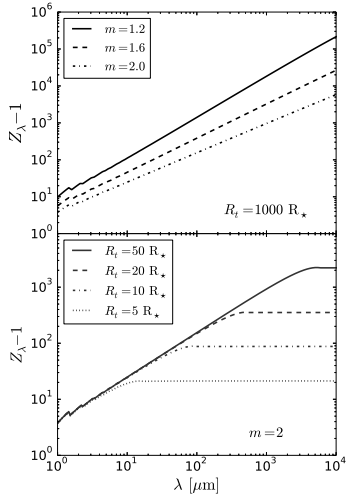}
\caption{Dependence of the excess flux ratio $Z_\lambda$ on the density exponent $m$ (top, $R_t=1000R_*$ fixed), and on the disc truncation radius $R_t$ (bottom, $m=2$ fixed). We adopted the reference stellar parameters, $\Sone=1.7$ g/cm$^2$ and a pole-on orientation in both cases.}
\label{f:flux_excess}
\end{figure}

The monochromatic excess flux ratio is given by
\begin{equation}Z_\lambda-1 = \frac{F_\lambda-F_*}{F_*},
\label{e:Zl}\end{equation}
where $F_\lambda$ is the total flux and $F_*$ is the photospheric flux. Figure~\ref{f:flux_excess} shows how $Z_\lambda$ varies with the wavelength for different values of $m$ and $R_t$. The disc emission is very sensitive to $m$. The slope of $Z_\lambda$ changes abruptly when $\Rbar\approx R_t$.
Such results suggest that the density exponent, as determined by detailed spectral energy distribution (SED) studies, can be used in order to indirectly detect otherwise unseen binary companions.

In binary systems with highly elliptical orbits ($\epsilon>0.6$), the disc inside the truncation region no longer has azimuthal symmetry (see Fig.~\ref{f:e09}). In such systems, there are two limiting cases from an observational perspective: 
(i) the case where $\Rbar \ll R_t$, and 
(ii) $\Rbar\ge R_t$. 
In the first case, as the pseudo-photosphere is almost not perturbed by the orbital motion, small or no flux modulations are expected.
In the second case, however, the pseudo-photosphere is modified by the tidal perturbation of the secondary, and thus important flux variations may arise. 
We emphasize that the two cases above are wavelength dependent, since $\Rbar$ grows with $\lambda$ approximately as \mbox{$\Rbar\propto\lambda^{2.2/(2m+1.5)}$} \citep{ViCa15}. In a given system, whilst case (i) likely holds for short wavelengths (visible to near-infrared), case (ii) may be seen for long wavelengths, depending on the disc density and the orbital parameters.
Therefore, variability due to binarity will be, in general, more easily detected in the infrared and radio domains than in the optical and the near-infrared, although some Be/X-ray binaries also exhibit brightening at periastron \citep{ScCU13}.

\begin{figure}\centering
  \includegraphics[scale=.7]{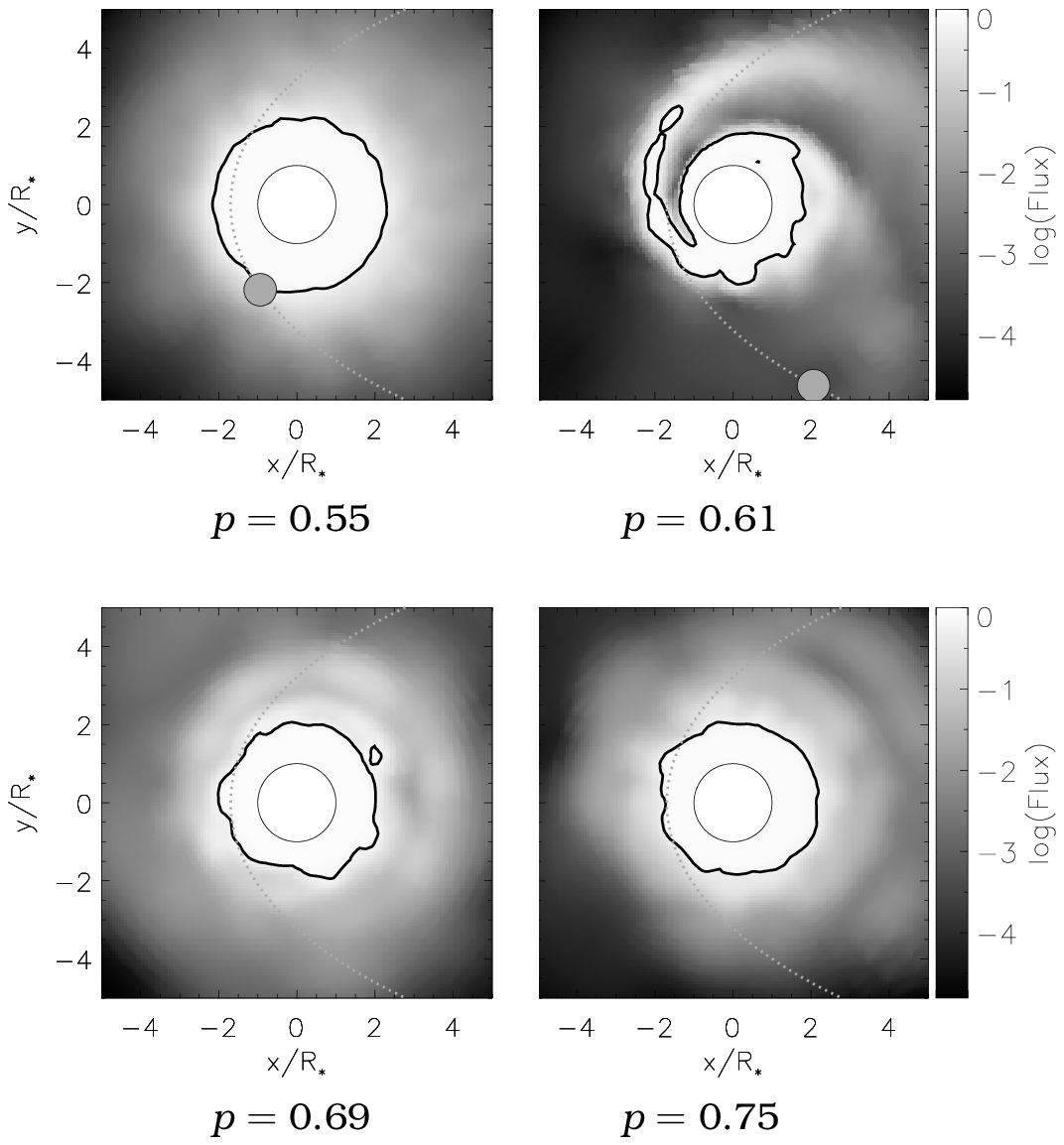}
\caption{Synthetic images at $3~\mu$m, for a system similar to the reference system but with $\epsilon=0.9$. The pseudo-photosphere shape (where $\tauobs=1$) is indicated by the solid curves, while the orbit of the secondary is represented by the dotted elliptical curves.}
\label{f:synthetic_img}
\interspace
  \includegraphics[scale=.7]{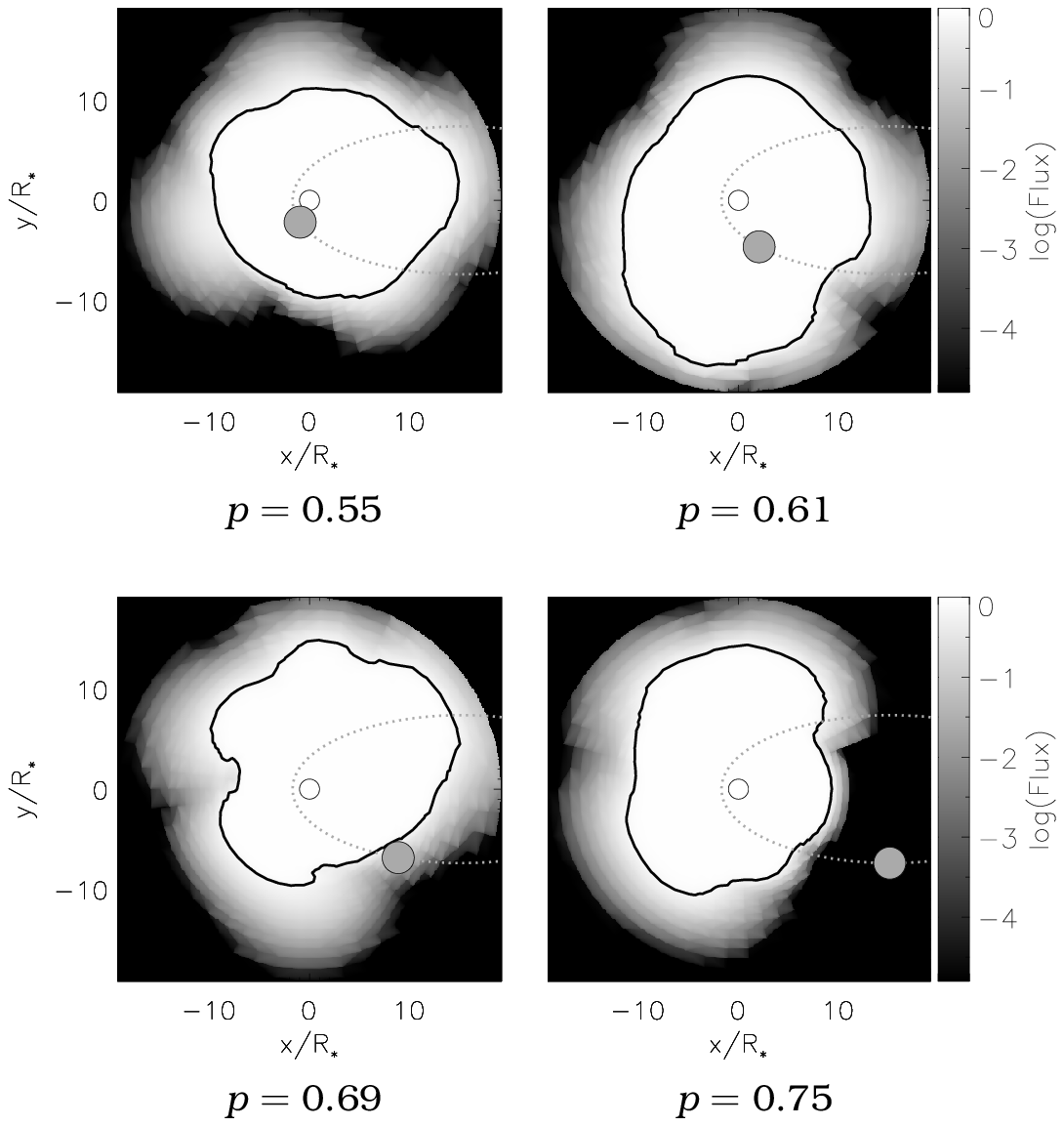}
 \caption{Same as Fig.~\ref{f:synthetic_img} but for $\lambda=1$ mm.}
 \label{f:sim-1mm}
\end{figure}

As an example, we computed the disc flux as a function of the orbital phase for the $\epsilon=0.9$ prograde case. We adopted the opacity of \citet{Bruss62} to compute $\tauobs$ at each disc position. A pole-on viewing angle was adopted.
To compute the total flux, we assumed a stellar blackbody emission of temperature $\Teff$ and a disc specific intensity
\begin{equation}\label{eq:spec_intens}
I_{\lambda}^{\mathrm{disc}}(r)=B_{\lambda}(\Td)[1-\exp(-\tauobs)],
\end{equation}
where $\Td=0.6\Teff$ is the isothermal disc temperature. The synthetic images at $3~\mu$m (Fig.~\ref{f:synthetic_img}) indeed show that $R_t$ is of the order of $\Rbar$, while at 1~mm (Fig.~\ref{f:sim-1mm}) global oscillations of the disc size determine the flux variation.

\begin{figure}\centering
\includegraphics[scale=.55]{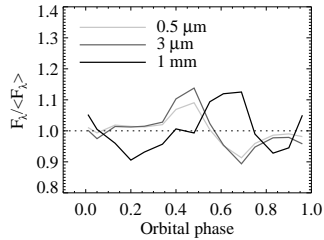}
\caption{Light curve along an orbital period at quasi-steady state for three wavelengths ($\lambda=0.5~\mu$m, $3~\mu$m, 1mm), normalised by its mean value over that period. The reference stellar parameters, with $\Sone=1.7$~g/cm$^2$, and pole-on orientation were adopted.}
\label{f:light_curve}
\end{figure}

Figure \ref{f:light_curve} shows the flux over an orbital period for three wavelengths, \mbox{$\lambda=0.5~\mu$m}, $3~\mu$m and 1~mm. The relative amplitude is similar for the three cases. However, unlike the long-wavelength curve, the phase-dependence of the short-wavelength light curves is similar, suggesting that these two wavelengths probe the emission from disc regions with a similar azimuthal structure.
This shows that it is possible to use a combination of short- and long-wavelength observations to study the tidal perturbations caused by a binary companion.

\section{Conclusions}\label{s:end}
As Be stars are often members of binary systems, it is expected that their discs interact in multiple ways with their companion.
In the past, \cite{OkBa02} studied the tidal interaction between the disc and the companion star in Be binaries and how it is affected by viscosity.
In this work, we built upon previous works, using the same {\threed} SPH code as the one used by \citeauthor{OkBa02}, in order to explore the density structure of the decretion disc more thoroughly, not only expanding to further ranges of a parameter space consisting of both disc (kinematic viscosity) and binary (orbital period, mass ratio, eccentricity) parameters, but also between azimuthally averaged and azimuthal-angle- and phase-dependent properties.

We examined the cases of circular and elliptical, prograde and retrograde orbits, for coplanar binaries, in which the orbital plane coincides with the equatorial plane of the disc.
Provided a sufficiently long evolution time that depends on all parameters (but is particularly long for high secondary-to-primary mass ratios, increases with increasing orbital period, and decreases with increasing eccentricity or viscosity parameter), the disc density profile relaxes in a quasi-steady state whose configuration generally depends on both the orbital phase and the azimuthal angle. The degree of dependence varies with the parameter values.

The most important effects are the truncation of the disc and the accumulation of matter inward of the truncation radius. The latter causes the disc density to drop less steeply with the radial distance than the case of an isolated Be star disc.
If a inner density fall-off exponent $m<2$ is measured for a disc known to be stable for a long time (a few years at least), the presence of a binary companion should at least be considered as a possibility.

With respect to the effect of eccentricity, two limiting cases were found. In the case of circular to moderately eccentric ($\epsilon\lesssim0.6$) orbits, the inner disc is relatively unaffected by either the phase or the azimuthal angle. However, the outer disc and the extent of the inner disc, quantified by the truncation radius, are azimuthally dependent. In general, in circular orbits the overall configuration does not change but rather only rotates with the orbital phase, following the orbit of the secondary.
In highly eccentric orbits the situation becomes more complex: the disc structure is strongly perturbed as the companion passes the periastron, forming spiral arms that are destroyed so much quicker as higher the eccentricity is.

High values of both the viscosity parameter and the orbital period are followed by an increase of the disc size, as well as a gradual disappearance of the azimuthal dependence, so that the disc becomes almost axisymmetric. On the contrary, the disc becomes smaller in size and more noticably elongated in high binary mass ratios; but its total mass is bigger and, constrained in a smaller disc extent, the accumulation effect is exhibited in a very high degree.

The only case where there was neither an orbital nor an azimuthal dependence was the retrograde circular system. In retrograde systems the time of interaction between the secondary and the disc particles is too low. In circular retrograde systems this results in that the only effect of the existence of a companion star is the truncation of the disc, simply because the companion passes through it.
The situation changes with increased eccentricity of retrograde systems, where the secondary largely destroys the disc after the periastron. Nonetheless, the disc reaches an almost circular but very sparse configuration much before the next apastron, which is maintained for the largest part of the orbital cycle.

Performing simulations for various values of the semi-major axis for circular systems, we concluded that there is no lower limit for the orbital separation in order that a decretion disc is formed. This means that the rarity of Be stars in closer orbits ($\Porb\lesssim30$ days) is a real phenomenon, and not a result of the absence of discs (in which case the star would no longer be a Be star) for close binaries.

Finally, we studied the effects of binarity in the disc continuum emission.
With the infrared and radio SED being sensitive to the extent and density slope of the disc, binarity causes measurable changes in the emergent spectrum. Furthermore, in the case of elliptical orbits, the disc brightness is expected to vary along the orbital cycle.
In both cases, infrared and radio studies of Be stars can be used as a tool to either detect unseen binaries via their effects on the disc or to further characterise known binaries.

In all simulations presented (except for the circular retrograde system), no matter the eccentricity or the direction of disc rotation, the Be decretion disc structure exhibits a periodicity, which is not present in isolated stars, therefore is intrinsic of the binary nature of the system.
Hence the tidal effects on the structure of the disc caused by the binary interaction may provide an explanation for the small-amplitude periodic $V/R$ variations in Be binaries. This hypothesis will be tested in a future publication, where detailed radiative transfer calculations for the simulations shown here will be conducted.

In general, the fact that our simulations show that the fall-off density exponent and the extent of the inner disc strongly depend on $\ass$, on the binary parameters $\Porb$, $q_r$, $\epsilon$, as well as on the orbital direction, suggests that disc studies can in principle be used to study unseen (and previously undetected) companions, using their effects on the disc \citep[see][]{KCR15}. This is particularly important in the case that the companion is a low-mass pre-main-sequence star, which is the most difficult to detect.

    \subsection*{Acknowledgements}
The authors thank the referee, V.~Icke, as well as D.~M.~F\~aes and R.~Klement, for their constructive comments that helped improve this manuscript.
      This work made use of the computing facilities of the Laboratory of Astroinformatics (IAG/USP, NAT/Unicsul), whose purchase was made possible by the Brazilian agency FAPESP (grant \mbox{2009/54006-4}) and the \mbox{INCT-A}. DP acknowledges support from FAPESP (grant \mbox{2013/16801-2}). ACC acknowledges support from CNPq (grant \mbox{307594/2015-7}) and FAPESP (grant \mbox{2015/17967-7}). RGV acknowledges the support from FAPESP (grant 2012/20364-4).
      CEJ acknowledges support from NSERC, the National Sciences and Research Council of Canada and the joint FAPESP/Western researcher exchange program.
\bibliography{biblio}
\end{document}